\documentclass[journal]{IEEEtran}
\IEEEoverridecommandlockouts
\usepackage{times,amsmath,color,amssymb,graphicx,epsfig,cite,psfrag,subfigure,balance}
\usepackage{svg}
\usepackage{caption}
\DeclareCaptionLabelFormat{tcomm}{Fig. \thefigure.}  
\usepackage{capt-of}
\usepackage{amsfonts,pifont,enumerate,cases}
\usepackage{mathrsfs} 
\usepackage[table]{xcolor} 
\usepackage{verbatim} 
\usepackage{bm}
\usepackage{cuted,stfloats}
\usepackage{algorithm}
\usepackage{algorithmic}

\allowdisplaybreaks
\usepackage{longtable}
\usepackage{blindtext}
\usepackage{multirow}
\usepackage{float}
\usepackage{threeparttable}
\usepackage{makecell}
\usepackage[utf8]{inputenc}
\usepackage{url}
\usepackage{booktabs}
\usepackage{amssymb}
\usepackage{bbding}
\usepackage{pifont}
\usepackage{wasysym}
\usepackage{utfsym}
\usepackage{fontawesome}
\usepackage[algo2e,ruled,linesnumbered,lined,boxed,commentsnumbered]{algorithm2e}
\usepackage[
    colorlinks=true,
    linkcolor=blue,
    citecolor=blue,
    urlcolor=magenta
]{hyperref}

\newcounter{MYtempeqncnt}

\begin{document}
\title{Beamforming and Trajectory Optimization for Aerial RIS-Aided Secure ISAC via Model Predictive Control}
\author{Zhendong Li, Zesheng Zhang, Zhou Su, Ziwei Liu, Ying Wang, and Wen Chen
\thanks{Zhendong Li and Zesheng Zhang are with the School of Information and Communication Engineering, Xi’an Jiaotong University, Xi’an 710049, China (email: lizhendong@xjtu.edu.cn; 2226215253@stu.xjtu.edu.cn). Zhou Su is with the School of Cyber Science and Engineering, Xi'an Jiaotong University, Xi'an 710049, China (email: zhousu@ieee.org). Ziwei Liu and Wen Chen are with the Department of Electronic Engineering, Shanghai Jiao Tong University, Shanghai 200240, China (e-mail: ziweiliu@sjtu.edu.cn; wenchen@sjtu.edu.cn). Ying Wang is with the State Key Laboratory of Networking and Switching Technology, Beijing University of Posts and Telecommunications, Beijing 100876, China (wangying@bupt.edu.cn).}\thanks{(Corresponding author: Zhou Su)}
\vspace{-1.5em}
}
\maketitle

\begin{abstract}
This paper proposes a novel aerial reconfigurable intelligent surface (RIS)-aided secure integrated sensing and communication (ISAC) architecture. An unmanned aerial vehicle (UAV)-mounted RIS establishes virtual line-of-sight links to bypass blockages, serving legitimate users while sensing potential eavesdroppers. To ensure communication security and sensing reliability, we formulate an infinite-horizon dynamic control problem maximizing the long-term average system secrecy rate and minimizing UAV flight energy. This formulation jointly optimizes the UAV trajectory, active beamforming with artificial noise (AN) at the base station, and RIS phase shifts, subject to strict radar signal-to-noise ratio and UAV kinematic constraints. Due to the highly coupled non-convex variables, solving this problem directly is intractable. Therefore, we propose an online joint optimization framework based on model predictive control (MPC). Within each receding horizon, alternating optimization is employed to decouple the problem into three subproblems, efficiently solved via successive convex approximation and difference of convex programming. Extensive simulations demonstrate that our MPC-based algorithm significantly outperforms static baselines in secrecy rate, exhibiting superior online correction capability and robustness against environmental perturbations.
\end{abstract}

\begin{IEEEkeywords}
Aerial RIS, secure ISAC, trajectory optimization, beamforming, model predictive control.
\end{IEEEkeywords}

\section{Introduction}
\IEEEPARstart{I}{ntegrated} sensing and communication (ISAC) is envisioned as a key enabler for 6G networks, with extensive research highlighting its spectral and hardware efficiency \cite{T1,T2,T3}. However, the dual-functional nature of ISAC poses severe physical-layer security challenges. Because sensing signals are inherently broadcasted over wide areas for target detection, they are highly susceptible to malicious eavesdropping. Furthermore, the sensing target itself may act as an untrusted node. This deep coupling of sensing and communication renders traditional security mechanisms inadequate, exposing the system to severe information leakage risks.

To counter these security threats, extensive literature has explored secure ISAC systems through beamforming design \cite{T4,T5,T6,bf1}, artificial noise (AN) strategies \cite{T8,T9,T11}, and performance bounds \cite{T12,T13}. Specifically, \cite{T4} proposed a robust beamforming scheme that maximizes the secrecy rate and minimizes the Cramér-Rao bound for target estimation. To combat covert eavesdropping, \cite{T5} integrated robust precoding with AN, while \cite{T6} investigated cell-free ISAC systems, jointly optimizing transceivers to prevent both communication leakage and sensing eavesdroppers from extracting target features. Furthermore, \cite{T8} utilized sensing signals as AN to mask communications, and \cite{T9} leveraged AN to protect the sensing mutual information of legitimate targets from illegal extraction.
However, most existing designs rely heavily on ideal line-of-sight (LoS) channels. In practical urban environments, these links are highly vulnerable to physical blockages by static obstacles and dynamic occlusions \cite{T14}. Such blockages force signals into severe non-line-of-sight (NLoS) propagation, drastically degrading both the ISAC coverage and the secrecy capacity of legitimate links \cite{T16}. To overcome this LoS bottleneck, deploying reconfigurable intelligent surfaces (RIS) to establish controllable virtual links has emerged as a critical paradigm for restoring and enhancing ISAC security and coverage.

RIS can proactively reshape the wireless environment to bypass physical blockages and enhance ISAC performance \cite{T17,T18,T19,RIS1}. Recent efforts have successfully integrated RIS into ISAC systems. For instance, \cite{T20} jointly optimized transmit waveforms and passive reflection to improve the radar signal-to-interference-plus-noise ratio (SINR). To overcome the double path loss of passive arrays, \cite{T21} maximized radar SINR using an active RIS architecture. Furthermore, \cite{T22} investigated an active simultaneously transmitting and reflecting RIS (STAR-RIS) for full-duplex ISAC, balancing multi-target sensing and communication sum-rate.
Nevertheless, fixed terrestrial RISs suffer from limited spatial degrees of freedom and remain highly susceptible to secondary blockages. Consequently, aerial RIS, with its high-altitude 3D mobility, has emerged to reconstruct high-quality LoS links \cite{T23}. While joint trajectory optimization and AN design for aerial RIS have enhanced security in traditional networks \cite{T24,T25}, these works focus solely on communication. Due to the deeply coupled dual functionalities in secure ISAC, fully exploiting aerial RIS mobility via joint trajectory planning and beamforming to mitigate asymmetric channel degradation and eavesdropping remains a critical open issue.

Furthermore, existing joint trajectory and beamforming designs predominantly rely on static offline frameworks \cite{T26,T27,T28,T29,T30}. For instance, alternating optimization (AO) and successive convex approximation (SCA) are widely used to tackle complex non-convex constraints \cite{T26,T28,T30}. However, these classical paradigms exhibit significant limitations in highly dynamic secure ISAC systems. They are inherently myopic and require unrealistic global a priori information over the entire flight. Lacking closed-loop feedback, these static algorithms fail to adapt to real-time variations, such as stochastic channels or dynamic eavesdropping threats, leading to severe performance degradation.
To break this impasse, model predictive control (MPC) emerges as a highly promising online optimization paradigm \cite{T31}. By making forward-looking decisions within a receding horizon, MPC effectively compensates for environmental uncertainties \cite{T32}. While MPC has demonstrated exceptional performance in UAV trajectory planning \cite{T33} and secure communications \cite{T34}, its application to aerial RIS-aided secure ISAC networks remains unexplored. Given the inherently coupled dual functionalities, how to leverage MPC to synergistically guarantee communication confidentiality and sensing precision via the joint dynamic control of trajectories and beamforming remains a pressing open problem. To bridge this critical gap, this paper proposes a novel MPC-based dynamic joint optimization framework. Unlike conventional static designs, the proposed framework organically integrates receding horizon control with low-complexity alternating optimization and difference of convex (DC) programming, proactively adapting to environmental uncertainties while efficiently resolving the spatio-temporal coupling.

Motivated by the aforementioned gaps, this paper investigates an aerial RIS-aided secure ISAC architecture, focusing on joint beamforming and trajectory optimization via MPC. By jointly optimizing the UAV trajectory, active beamforming with AN at the base station, and passive RIS phase shifts, we aim to maximize the long-term average system secrecy rate and minimize the UAV flight energy consumption, while strictly guaranteeing the radar sensing performance. The main contributions of this paper are summarized as follows:
\begin{itemize}
    \item We propose a novel aerial RIS-aided secure ISAC architecture to combat blocked direct links and eavesdropping threats. Distinct from conventional static offline designs, we formulate an MPC-based infinite-horizon dynamic control problem. Subject to stringent radar signal-to-noise ratio (SNR) thresholds and UAV kinematic constraints, this formulation jointly optimizes the 3D trajectory, transceivers, AN, and RIS phase shifts to maximize the long-term average secrecy rate and minimize flight energy consumption.
    
    \item To tackle the highly non-convex and spatio-temporally coupled problem, we propose a receding horizon online framework. At each step, AO framework is utilized to decouple the problem into three tractable subproblems. Specifically, SCA handles the non-convex trajectory and rate constraints, while semidefinite relaxation (SDR) paired with a DC programming-based penalty method efficiently resolves the non-convex rank-1 constraints of the beamforming and phase shift matrices.
    
    \item Extensive simulations demonstrate the advantages of aerial RIS mobility in evading eavesdroppers and reconstructing secure links. Compared to static baselines, our proposed online MPC algorithm achieves significant gains in secrecy rate and sensing performance. Furthermore, it exhibits exceptional online trajectory correction capabilities and system robustness against unexpected environmental perturbations, such as continuous crosswinds.
\end{itemize}

\textit{Notations:} Lower-case, bold lower-case, and bold upper-case letters denote scalars, vectors, and matrices, respectively. $|a|$ is the modulus of scalar $a$. $\|\mathbf{a}\|$ and $[\mathbf{a}]_i$ are the Euclidean norm and $i$-th element of vector $\mathbf{a}$. $\mathrm{diag}(\mathbf{a})$ denotes a diagonal matrix with elements of $\mathbf{a}$ on its main diagonal. For matrix $\mathbf{A}$, $\mathbf{A}^T$, $\mathbf{A}^H$, $\mathrm{Tr}(\mathbf{A})$, and $\mathrm{rank}(\mathbf{A})$ represent its transpose, conjugate transpose, trace, and rank. $[\mathbf{A}]_{i,j}$ is the $(i,j)$-th entry of $\mathbf{A}$. $\otimes$ and $\odot$ denote the Kronecker and Hadamard products. $\langle \cdot, \cdot \rangle$ represents the inner product. $\|\mathbf{A}\|_2$ and $\partial\|\mathbf{A}\|_2$ denote the spectral norm of $\mathbf{A}$ and its subgradient. $\mathbf{A} \succeq \mathbf{0}$ indicates a positive semidefinite matrix. $\mathbb{C}^{a \times b}$ is the space of $a \times b$ complex matrices. $\mathcal{CN}(\mathbf{0}, \mathbf{\Sigma})$ represents the circularly symmetric complex Gaussian distribution, and $\mathbb{E}[\cdot]$ denotes the statistical expectation.

		\section{System Model and Problem Formulation}\label{sec2}
\begin{figure}[htbp]
    \centering
    \includegraphics[height=45mm, width=86mm, keepaspectratio]{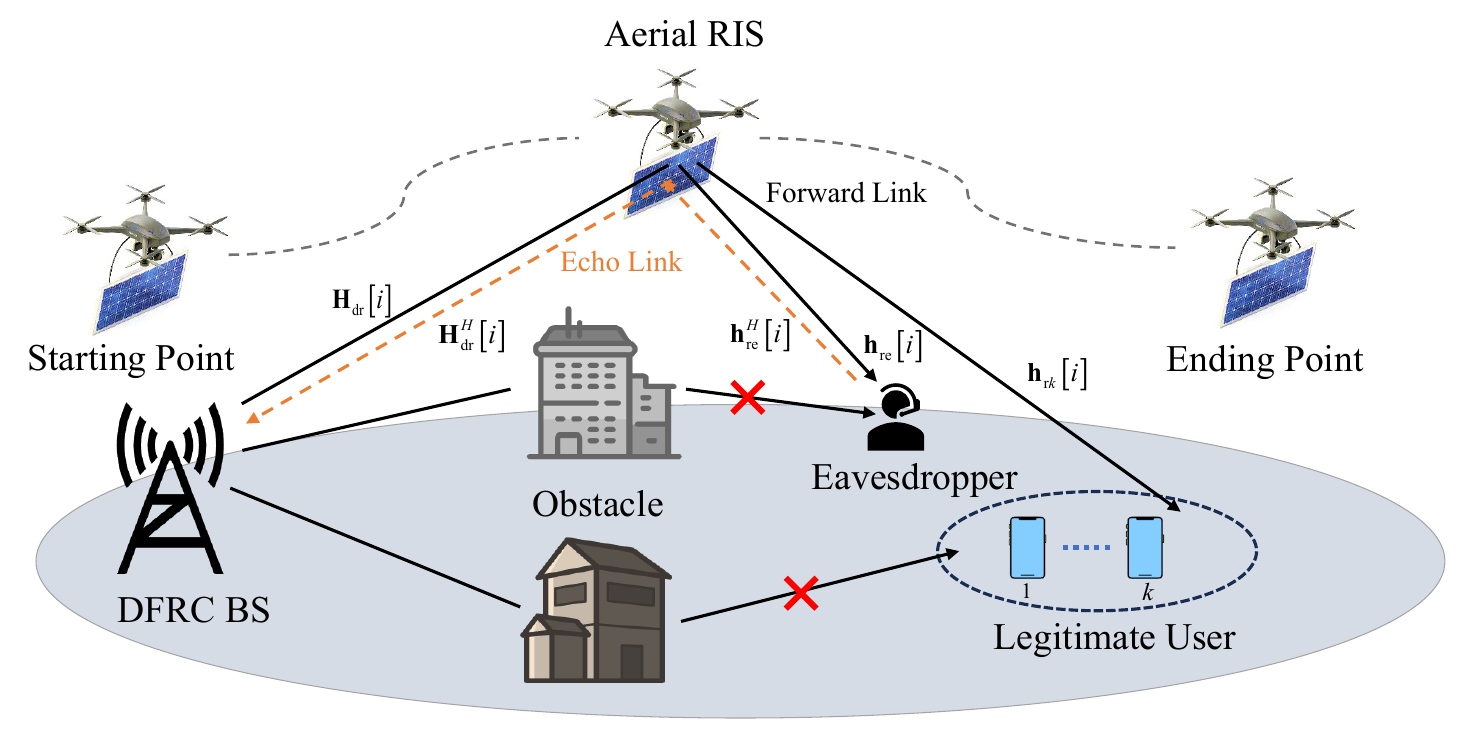}
    \caption{Illustration of the aerial RIS-aided secure ISAC system.}
    \label{fig1}
    \vspace{2em}
\end{figure}

As depicted in Fig.~\ref{fig1}, we consider an aerial RIS-aided secure ISAC network. In this system, a dual-functional radar and communication (DFRC) base station (BS) equipped with an $N_t$-transmit and $N_r$-receive uniform linear array (ULA) simultaneously communicates with $K$ single-antenna legitimate users and performs radar sensing to track a single-antenna potential eavesdropper. We assume that the direct communication links between the BS and the users are severely obstructed. To overcome this, a UAV equipped with a uniform planar array (UPA) of $M = M_x \times M_y$ reflecting elements is deployed as an aerial RIS. This deployment flexibly reconstructs the signal paths and extends the overall network coverage.
Regarding the signal transmission mechanism, the RIS reflecting elements operate in the full-reflection mode (i.e., with unit reflection amplitude). To preclude hardware conflicts and self-interference, a time-division protocol is adopted. Specifically, the entire operation within each time slot is temporally partitioned into a downlink transmission phase (optimizing for secure communication) and an uplink radar sensing phase (optimizing for echo reception). Accordingly, the corresponding phase-shift matrices are designed as follows:
\begin{align}
    \mathbf{\Theta} &= \text{diag} \left\{ e^{j\alpha_1}, \dots, e^{j\alpha_m}, \dots, e^{j\alpha_M} \right\}, \label{eq:theta} \\
    \mathbf{\Psi} &= \text{diag} \left\{ e^{j\beta_1}, \dots, e^{j\beta_m}, \dots, e^{j\beta_M} \right\}, \label{eq:psi}
\end{align}
where $\alpha_m, \beta_m \in [0, 2\pi)$ denote the phase shifts of the $m$-th reflecting element during the transmission and sensing phases, respectively. These phase shifts strictly satisfy the unit modulus constraint, i.e., $|e^{j\alpha_m}| = |e^{j\beta_m}| = 1, \forall m \in \{1, \dots, M\}$. Assuming fixed time allocations for both phases, the constant time-splitting factor is omitted in subsequent rate and SNR expressions for brevity, which thus represent instantaneous performance.

\subsection{UAV Mobility Model}
We assume that the locations of the DFRC BS, the $k$-th legitimate user, and the potential eavesdropper are stationary, which are denoted by ${{\bf{q}}_0} = {\left[ {{x_0},{y_0},0} \right]^T}$, ${{\bf{q}}_k} = {\left[ {{x_k},{y_k},0} \right]^T}$, and ${{\bf{q}}_{\rm{e}}} = {\left[ {{x_{\rm{e}}},{y_{\rm{e}}},0} \right]^T}$. We focus on the dynamic trajectory of the UAV. Consider a rotary-wing UAV carrying the RIS, which flies from a predetermined starting point $\mathbf{q}_{\mathrm{A}} = [x_{\mathrm{A}}, y_{\mathrm{A}}, z_{\mathrm{A}}]^T \in \mathbb{R}^3$ to a specified destination $\mathbf{q}_{\mathrm{B}} = [x_{\mathrm{B}}, y_{\mathrm{B}}, z_{\mathrm{B}}]^T \in \mathbb{R}^3$. The entire flight period is partitioned into $I$ time slots of equal duration $t_c$, indexed by $i = 0, 1, \dots, I-1$. During the flight, the 3D mobility model and physical constraints of the UAV are described as:
\begin{subequations} \label{eq:uav_mobility}
\begin{align}
    \mathbf{q}[0] &= \mathbf{q}_{\mathrm{A}}, \quad \mathbf{q}[i+1] = \mathbf{q}[i] + \mathbf{v}[i]t_c, \label{eq:pos_update} \\
    \| (v_x[i], v_y[i]) \| &\le V_{\max}, \quad v_z[i] \le U_{\max}, \label{eq:vel_limits} \\
    Z_{\min} \le q_z[i] &\le Z_{\max}, \quad \| \mathbf{v}[i+1] - \mathbf{v}[i] \| \le a_{\max}t_c. \label{eq:alt_accel_limits}
\end{align}
\end{subequations}
where $\mathbf{q}[i] = [q_x[i], q_y[i], q_z[i]]^T$ denotes the position vector of the UAV in time slot $i$, with $Z_{\min}$ and $Z_{\max}$ representing the minimum and maximum flight altitude limits, respectively. Similarly, $\mathbf{v}[i] = [v_x[i], v_y[i], v_z[i]]^T$ represents the velocity vector in time slot $i$. Here, the horizontal velocity $\mathbf{v}_h[i] = [v_x[i], v_y[i]]^T$ is bounded by $V_{\max}$, and the vertical velocity $v_z[i]$ is limited by $U_{\max}$. Additionally, $a_{\max}$ and $t_c$ denote the maximum acceleration of the UAV and the duration of each time slot, respectively. In time slot $i$, the total power consumption of the UAV can be modeled as\cite{T33}:
\begin{equation}
\begin{split}
    P(\mathbf{v}[i]) = & \frac{W^2}{\sqrt{2}\rho S} \left( \|\mathbf{v}_h[i]\|^2 + \sqrt{\|\mathbf{v}_h[i]\|^4 + 4V_h^4} \right)^{-\frac{1}{2}} \\
    & + W v_z[i] + \frac{\zeta \rho S}{8} \|\mathbf{v}_h[i]\|^3,
\end{split}
\label{eq:power_model}
\end{equation}
where $W$ is the total mass of the aerial RIS platform, $\rho$ denotes the air density, and $S$ represents the area of the UAV rotor disk. Additionally, $V_h = \sqrt{W / (2\rho S)}$ is the hovering parameter and $\zeta$ denotes the profile drag coefficient. In \eqref{eq:power_model}, the first term represents the horizontal flight power, the second term denotes the power consumption for vertical flight (ascent or descent), and the third term corresponds to the rotor profile drag power.
\subsection{Channel Model}
We assume that the direct communication links between the DFRC BS and both the legitimate user and the eavesdropper are severely obstructed. Consequently, an aerial RIS is deployed to assist in providing secure communication services to the legitimate user while simultaneously sensing the eavesdropper. Given the high-altitude deployment of the UAV and the surrounding scattering environment, all channels associated with the aerial RIS are modeled as Rician fading channels.

Since the DFRC BS is equipped with a  ULA, and taking the mobility of the aerial RIS into account, the array response vector at the $i$-th time slot can be expressed as:
\begin{equation}
    \mathbf{a}_{\mathrm{ULA}}(\theta[i]) = \left[ 1, \dots, e^{j\frac{2\pi}{\lambda}(N-1)d_1 \sin(\theta[i])} \right]^T \in \mathbb{C}^{N \times 1},
\end{equation}
where $\theta[i]$ denotes the elevation angle from the DFRC BS to the aerial RIS at the $i$-th time slot. The parameter $d_1$ represents the element spacing of the DFRC BS, which is typically set to $\lambda/2$, with $\lambda$ being the wavelength of the carrier signal.
As the aerial RIS employs a UPA, it possesses both horizontal and vertical array responses. Accounting for the mobility of the aerial RIS, the corresponding response vectors $\mathbf{a}_{\mathrm{horiz}} \in \mathbb{C}^{M \times 1}$ and $\mathbf{a}_{\mathrm{vert}} \in \mathbb{C}^{M \times 1}$ at the $i$-th time slot are expressed as:
\begin{equation}
    \mathbf{a}_{\mathrm{horiz}}\left(\theta_{\mathrm{t}}[i], \phi_{\mathrm{t}}[i]\right) = \left[ 1, \dots, e^{j(M-1)\frac{2\pi}{\lambda}d_2 \sin(\phi_{\mathrm{t}}[i]) \cos(\theta_{\mathrm{t}}[i])} \right]^T,
\end{equation}
\begin{equation}
    \mathbf{a}_{\mathrm{vert}}\left(\theta_{\mathrm{t}}[i]\right) = \left[ 1, \dots, e^{j(M-1)\frac{2\pi}{\lambda}d_3 \sin(\theta_{\mathrm{t}}[i])} \right]^T,
\end{equation}
where $d_2$ and $d_3$ denote the horizontal and vertical element spacings of the RIS, respectively, which are also typically set to $\lambda/2$. Furthermore, $\theta_{\mathrm{t}}[i]$ and $\phi_{\mathrm{t}}[i]$ represent the elevation and azimuth angles from the aerial RIS to the target at the $i$-th time slot, respectively.Consequently, the overall array response vector of the aerial RIS at the $i$-th time slot is given by the Kronecker product of its horizontal and vertical components:
\begin{equation}
    \mathbf{a}_{\mathrm{RIS}}\left[\theta_{\mathrm{t}}[i], \phi_{\mathrm{t}}[i]\right] = \mathbf{a}_{\mathrm{horiz}}\left(\theta_{\mathrm{t}}[i], \phi_{\mathrm{t}}[i]\right) \otimes \mathbf{a}_{\mathrm{vert}}\left(\theta_{\mathrm{t}}[i]\right).
\end{equation}

Accordingly, the LoS channel matrix $\mathbf{H}_{\mathrm{dr}}^{\mathrm{LoS}}[i] \in \mathbb{C}^{N \times M}$ from the DFRC BS to the aerial RIS during the $i$-th time slot can be modeled as:
\begin{equation}
    \mathbf{H}_{\mathrm{dr}}^{\mathrm{LoS}}[i] = \mathbf{a}_{\mathrm{ULA}}\left(\theta_{\mathrm{Ds}}[i]\right) \mathbf{a}_{\mathrm{RIS}}^{H}\left[\theta_{\mathrm{Ds}}[i], \phi_{\mathrm{Ds}}[i]\right],
\end{equation}
where $\theta_{\mathrm{Ds}}[i]$ denotes the elevation angle of departure (AoD) at the DFRC BS, and $\phi_{\mathrm{Ds}}[i]$ represents the azimuth AoD from the DFRC BS to the aerial RIS.

Similarly, the LoS channel vector $\mathbf{h}_{\mathrm{rk}}^{\mathrm{LoS}}[i] \in \mathbb{C}^{M \times 1}$ from the aerial RIS to the $k$-th legitimate user in the $i$-th time slot is expressed as:
\begin{equation}
    \mathbf{h}_{\mathrm{r}k}^{\mathrm{LoS}}[i] = \mathbf{a}_{\mathrm{horiz}}\left(\theta_{k}[i], \phi_{k}[i]\right) \otimes \mathbf{a}_{\mathrm{vert}}\left(\theta_{k}[i]\right),
\end{equation}
where $\theta_{k}[i]$ and $\phi_{k}[i]$ denote the elevation and azimuth angles from the aerial RIS to the $k$-th legitimate user, respectively. Furthermore, the LoS channel vector $\mathbf{h}_{\mathrm{re}}^{\mathrm{LoS}}[i] \in \mathbb{C}^{M \times 1}$ from the aerial RIS to the potential eavesdropper is given by:
\begin{equation}
    \mathbf{h}_{\mathrm{re}}^{\mathrm{LoS}}[i] = \mathbf{a}_{\mathrm{horiz}}\left(\theta_{\mathrm{e}}[i], \phi_{\mathrm{e}}[i]\right) \otimes \mathbf{a}_{\mathrm{vert}}\left(\theta_{\mathrm{e}}[i]\right),
\end{equation}
where $\theta_{\mathrm{e}}[i]$ and $\phi_{\mathrm{e}}[i]$ represent the elevation and azimuth angles from the aerial RIS to the eavesdropper. It is worth noting that these angles are acquired through the sensing capabilities of the DFRC BS.
\begin{figure*}[!t]
\normalsize
\setcounter{MYtempeqncnt}{\value{equation}}
\setcounter{equation}{20} 

\begin{equation} \label{eq:achievable_rate}
    R_k[i] = \log_2 \left( 1 + \frac{\left| \mathbf{h}_{\mathrm{r}k}^H[i] \mathbf{\Theta}[i] \mathbf{H}_{\mathrm{dr}}^H[i] \mathbf{w}_k[i] \right|^2}{ \sum\limits_{c \neq k}^K \left| \mathbf{h}_{\mathrm{r}k}^H[i] \mathbf{\Theta}[i] \mathbf{H}_{\mathrm{dr}}^H[i] \mathbf{w}_c[i] \right|^2 + \left| \mathbf{h}_{\mathrm{r}k}^H[i] \mathbf{\Theta}[i] \mathbf{H}_{\mathrm{dr}}^H[i] \mathbf{m}[i] \right|^2 + \sigma_k^2[i] } \right).
\end{equation}

\setcounter{equation}{\value{MYtempeqncnt}}
\hrulefill
\vspace*{4pt}
\end{figure*}

Based on the Rician fading assumption, the overall channel matrix between the DFRC BS and the aerial RIS at the $i$-th time slot is formulated as:
\begin{equation}
    \bar{\mathbf{H}}_{\mathrm{dr}}[i] = \sqrt{\frac{\kappa_{\mathrm{dr}}}{1+\kappa_{\mathrm{dr}}}} \mathbf{H}_{\mathrm{dr}}^{\mathrm{LoS}}[i] + \sqrt{\frac{1}{1+\kappa_{\mathrm{dr}}}} \mathbf{H}_{\mathrm{dr}}^{\mathrm{NLoS}}[i],
\end{equation}
where the elements of the NLoS component follow a standard complex Gaussian distribution, i.e., $[\mathbf{H}_{\mathrm{dr}}^{\mathrm{NLoS}}[i]]_{n,m} \sim \mathcal{CN}(0,1)$, and $\kappa_{\mathrm{dr}}$ denotes the Rician factor of the channel.
In the $i$-th time slot, the channel vector between the aerial RIS and the $k$-th legitimate user is modeled as:
\begin{equation}
    \bar{\mathbf{h}}_{\mathrm{r}k}[i] = \sqrt{\frac{\kappa_{\mathrm{r}k}}{1+\kappa_{\mathrm{r}k}}} \mathbf{h}_{\mathrm{r}k}^{\mathrm{LoS}}[i] + \sqrt{\frac{1}{1+\kappa_{\mathrm{r}k}}} \mathbf{h}_{\mathrm{r}k}^{\mathrm{NLoS}}[i], \quad \forall k,
\end{equation}
where the NLoS component follows $[\mathbf{h}_{\mathrm{rk}}^{\mathrm{NLoS}}[i]]_m \sim \mathcal{CN}(0,1)$, and $\kappa_{\mathrm{rk}}$ represents the Rician factor of this channel.Similarly, the channel vector from the aerial RIS to the eavesdropper in the $i$-th time slot is given by:
\begin{equation}
    \bar{\mathbf{h}}_{\mathrm{re}}[i] = \sqrt{\frac{\kappa_{\mathrm{re}}}{1+\kappa_{\mathrm{re}}}} \mathbf{h}_{\mathrm{re}}^{\mathrm{LoS}}[i] + \sqrt{\frac{1}{1+\kappa_{\mathrm{re}}}} \mathbf{h}_{\mathrm{re}}^{\mathrm{NLoS}}[i],
\end{equation}
where $[\mathbf{h}_{\mathrm{re}}^{\mathrm{NLoS}}[i]]_m \sim \mathcal{CN}(0,1)$, and $\kappa_{\mathrm{re}}$ denotes the corresponding Rician factor.

Incorporating large-scale fading, the aggregate channel matrix $\mathbf{H}_{\mathrm{dr}}[i] \in \mathbb{C}^{N \times M}$ (DFRC BS to aerial RIS), the channel vector $\mathbf{h}_{\mathrm{r}k}[i] \in \mathbb{C}^{M \times 1}$ (aerial RIS to $k$-th user), and the channel vector $\mathbf{h}_{\mathrm{re}}[i] \in \mathbb{C}^{M \times 1}$ (aerial RIS to eavesdropper) in the $i$-th time slot are expressed as:
\begin{subequations} \label{eq:channel_models}
\begin{align}
    \mathbf{H}_{\mathrm{dr}}[i] &= \sqrt{\rho_0 \left( \frac{\|\mathbf{q}[i] - \mathbf{q}_0\|}{d_0} \right)^{-\alpha_{\mathrm{dr}}}} \bar{\mathbf{H}}_{\mathrm{dr}}[i], \label{eq:H_dr} \\
    \mathbf{h}_{\mathrm{r}k}[i] &= \sqrt{\rho_0 \left( \frac{\|\mathbf{q}[i] - \mathbf{q}_k\|}{d_0} \right)^{-\alpha_{\mathrm{r}k}}} \bar{\mathbf{h}}_{\mathrm{r}k}[i], \label{eq:h_rk} \\
    \mathbf{h}_{\mathrm{re}}[i] &= \sqrt{\rho_0 \left( \frac{\|\mathbf{q}[i] - \mathbf{q}_{\mathrm{e}}\|}{d_0} \right)^{-\alpha_{\mathrm{re}}}} \bar{\mathbf{h}}_{\mathrm{re}}[i]. \label{eq:h_re}
\end{align}
\end{subequations}
where $\rho_0$ denotes the reference path loss at a distance of $d_0$ (with $d_0 = 1\,\text{m}$), and $\alpha_{\mathrm{dr}}$, $\alpha_{\mathrm{rk}}$, and $\alpha_{\mathrm{re}}$ represent the path loss exponents for the respective channels.
\subsection{Signal Model and Performance Metrics}
In the $i$-th time slot, the DFRC BS serves a set of $K$ legitimate users, denoted by $\mathcal{K}$, via the aerial RIS. To further enhance secure communication performance, AN is injected at the DFRC BS. Consequently, the transmitted signal vector at the DFRC BS in the $i$-th time slot is modeled as:
\begin{equation}
    \mathbf{x}[i] = \sum_{k \in \mathcal{K}} \mathbf{w}_k[i] b_k[i] + \mathbf{m}[i],
\end{equation}
where $\mathbf{w}_k[i] \in \mathbb{C}^{N \times 1}$ represents the beamforming vector for the $k$-th user. The scalar $b_k[i] \in \mathbb{C}$ denotes the information-bearing symbol for the $k$-th legitimate user, satisfying the unit power constraint $\mathbb{E}\{|b_k[i]|^2\} = 1, \forall k$. The vector $\mathbf{m}[i] \in \mathbb{C}^{N \times 1}$ represents the transmitted AN signal, which follows a circularly symmetric complex Gaussian (CSCG) distribution, i.e., $\mathbf{m}[i] \sim \mathcal{CN}(\mathbf{0}, \mathbf{M}[i])$, with $\mathbf{M}[i] \succeq \mathbf{0}$ being the positive semi-definite covariance matrix.
Accordingly, the signal received by the $k$-th legitimate user in the $i$-th time slot can be expressed as:
\begin{equation}
\small 
\begin{split}
    y_k[i] =& \, \mathbf{h}_{\mathrm{r}k}^H[i] \mathbf{\Theta}[i] \mathbf{H}_{\mathrm{dr}}^H[i] \mathbf{w}_k[i] b_k[i] 
    + \sum_{c \neq k}^{K} \mathbf{h}_{\mathrm{r}k}^H[i] \mathbf{\Theta}[i] \mathbf{H}_{\mathrm{dr}}^H[i] \mathbf{w}_c[i] b_c[i] \\
    & + \mathbf{h}_{\mathrm{r}k}^H[i] \mathbf{\Theta}[i] \mathbf{H}_{\mathrm{dr}}^H[i] \mathbf{m}[i] + n_k[i],
\end{split}
\end{equation}
where the first term represents the desired signal for the target user, the second term denotes the multi-user interference, and the third term accounts for the interference caused by the AN. Finally, $n_k[i] \sim \mathcal{CN}(0, \sigma_k^2[i])$ represents the additive white Gaussian noise (AWGN) introduced at the receiver of the $k$-th legitimate user in the $i$-th time slot. Similarly, the signal intercepted by the potential eavesdropper in the $i$-th time slot can be expressed as:
\begin{equation}
\begin{split}
    y_{\mathrm{e}}[i] = & \sum_{k=1}^{K} \mathbf{h}_{\mathrm{re}}^H[i] \mathbf{\Theta}[i] \mathbf{H}_{\mathrm{dr}}^H[i] \mathbf{w}_k[i] b_k[i] \\
    & + \mathbf{h}_{\mathrm{re}}^H[i] \mathbf{\Theta}[i] \mathbf{H}_{\mathrm{dr}}^H[i] \mathbf{m}[i] + n_{\mathrm{e}}[i],
\end{split}
\end{equation}
where the first term denotes the intercepted information signals, the second term represents the interference caused by the AN, and $n_{\mathrm{e}}[i] \sim \mathcal{CN}(0, \sigma_{\mathrm{e}}^2[i])$ signifies the AWGN introduced at the eavesdropper's receiver in the $i$-th time slot.
Consequently, the echo signal reflected by the eavesdropper and received at the DFRC BS is modeled as:
\begin{equation}
    \mathbf{y}_{\mathrm{r}}[i] = \sum_{k=1}^{K} \mathbf{H}[i] \mathbf{w}_k[i] b_k[i] + \mathbf{H}[i] \mathbf{m}[i] + n_{\mathrm{r}}[i],
\end{equation}
where $\mathbf{H}[i] = \mathbf{H}_{\mathrm{dr}}[i] \mathbf{\Theta}[i] \mathbf{h}_{\mathrm{re}}[i] \mathbf{h}_{\mathrm{re}}^H[i] \mathbf{\Psi}[i] \mathbf{H}_{\mathrm{dr}}^H[i] \in \mathbb{C}^{N \times N}$ represents the equivalent round-trip sensing channel matrix. In this expression, the first term constitutes the useful signal for sensing, the second term corresponds to the AN signal, and $n_{\mathrm{r}}[i] \sim \mathcal{CN}(\mathbf{0}, \sigma_{\mathrm{r}}^2[i])$ denotes the AWGN at the DFRC BS antennas.
The SNR serves as a pivotal metric for evaluating radar sensing performance. Therefore, we adopt the radar SNR to quantify the system's sensing capability. In the $i$-th time slot, the radar SNR is formulated as:
\begin{equation}
    \text{SNR}^r[i] = \frac{\sum\limits_{k=1}^{K} \left| \mathbf{H}[i] \mathbf{w}_k[i] b_k[i] \right|^2 + \left| \mathbf{H}[i] \mathbf{m}[i] \right|^2}{\sigma_{\mathrm{r}}^2[i]}.
\end{equation}

The secrecy rate for the communication between the DFRC BS and the legitimate user is determined by the difference between the achievable rate of the legitimate link and the wiretapping rate of the eavesdropping link. Based on Shannon's theory, the achievable rate (in bps/Hz) for the $k$-th legitimate user in the $i$-th time slot is formulated as \eqref{eq:achievable_rate}.
To guarantee robust security, we consider a worst-case scenario where the potential eavesdropper possesses strong decoding capabilities, allowing it to cancel out multi-user interference from other legitimate users before decoding the target user's message. Consequently, the eavesdropping rate (in bps/Hz) of the potential eavesdropper for the $k$-th legitimate user in the $i$-th time slot is expressed as:
\addtocounter{equation}{1}
\begin{equation} \label{eq:next_equation_22}
    C_k[i] = \log_2 \left( 1 + \frac{\left| \mathbf{h}_{\mathrm{re}}^H[i] \mathbf{\Theta}[i] \mathbf{H}_{\mathrm{dr}}^H[i] \mathbf{w}_k[i] \right|^2}{ \left| \mathbf{h}_{\mathrm{re}}^H[i] \mathbf{\Theta}[i] \mathbf{H}_{\mathrm{dr}}^H[i] \mathbf{m}[i] \right|^2 + \sigma_{\mathrm{e}}^2[i] } \right).
\end{equation}
Therefore, the achievable secrecy rate for the $k$-th legitimate user in the $i$-th time slot is given by:
\begin{equation}
    D_k[i] = \left[ R_k[i] - C_k[i] \right]^+,
\end{equation}
where $[x]^+ = \max\{0, x\}$.
\subsection{MPC-Based Problem Formulation}
By jointly optimizing the UAV trajectory, active beamforming with AN, and RIS phase shifts, we can effectively enhance the secure ISAC performance. Here, we formulate an infinite-horizon control problem where the UAV navigates from $\mathbf{q}_{\mathrm{A}} = [x_{\mathrm{A}}, y_{\mathrm{A}}, z_{\mathrm{A}}]^T \in \mathbb{R}^3$ to $\mathbf{q}_{\mathrm{B}} = [x_{\mathrm{B}}, y_{\mathrm{B}}, z_{\mathrm{B}}]^T \in \mathbb{R}^3$. Subject to sensing, power, and kinematic constraints, our objective is to maximize the secrecy rate while minimizing UAV flight energy. The optimization problem (P1) is formulated as follows:
\begin{subequations} \label{prob:P1}
\begin{align}
    (\text{P1}) \quad & \max_{ \substack{\mathbf{q}[i], \mathbf{v}[i], \mathbf{w}_k[i], \\ \mathbf{m}[i], \mathbf{\Theta}[i], \mathbf{\Psi}[i]} } \quad \chi_1 \sum_{k \in \mathcal{K}} D_k[i] - \chi_2 \sum \|\mathbf{q}[i] - \mathbf{q}_B\|^2 \nonumber \\
    & \qquad \qquad - \chi_3 P\big(\mathbf{v}[i]\big), \nonumber \\
    \textrm{s.t.} \quad & \text{SNR}^r[i] \geq \gamma^r, \label{eq:p1_snr} \\
    & \mathbf{q}[i+1] = \mathbf{q}[i] + \mathbf{v}[i] t_c, \label{eq:p1_kinematics} \\
    & \left\| (v_x(i), v_y(i)) \right\| \leq V_{\max}, \label{eq:p1_v_hor} \\
    & v_z(i) \leq U_{\max}, \label{eq:p1_v_ver} \\
    & \left\| \mathbf{v}[i+1] - \mathbf{v}[i] \right\| \leq a_{\max} t_c, \label{eq:p1_acc} \\
    & Z_{\min} \leq q_z(i) \leq Z_{\max}, \label{eq:p1_height} \\
    & \sum_{k=1}^{K} \|\mathbf{w}_k[i]\|^2 + \|\mathbf{m}[i]\|^2 \leq P_{\max}, \label{eq:p1_power} \\
    & \left| [\mathbf{\Theta}[i]]_{m,m} \right| = 1, \label{eq:p1_ris_theta} \\
    & \left| [\mathbf{\Psi}[i]]_{m,m} \right| = 1. \label{eq:p1_ris_psi}
\end{align}
\end{subequations}
The objective function consists of three components: maximizing the sum secrecy rate of legitimate users, penalizing the distance to the destination to encourage swift arrival, and minimizing the UAV's propulsion energy. Regarding the constraints, \eqref{eq:p1_snr} ensures reliable eavesdropper tracking by enforcing a minimum radar SNR threshold $\gamma^r$. Constraints \eqref{eq:p1_kinematics}-\eqref{eq:p1_height} dictate the UAV's motion and spatial limits, ensuring state continuity \eqref{eq:p1_kinematics}, restricting horizontal/vertical velocities and accelerations for flight safety \eqref{eq:p1_v_hor}-\eqref{eq:p1_acc}, and confining the altitude within an optimal visibility range \eqref{eq:p1_height}. Furthermore, \eqref{eq:p1_power} imposes a maximum transmit power budget at the DFRC BS, while \eqref{eq:p1_ris_theta} and \eqref{eq:p1_ris_psi} enforce the unit modulus constraints on the RIS phase shift matrices for the downlink communication and uplink sensing phases, respectively.

Having established the aerial RIS-aided secure ISAC system model over an infinite time horizon, we must address the inherent challenges of dimensionality and prohibitive computational complexity associated with infinite-horizon optimization. To circumvent these issues, we propose an online joint optimization framework based on MPC. By leveraging receding horizon optimization and a closed-loop feedback mechanism, the original infinite-horizon problem is effectively decoupled into a sequence of finite-horizon subproblems. This strategy not only mitigates computational intractability but also significantly enhances the system's robustness against time-varying channel conditions and external disturbances.
Let $N_p$ denote the prediction horizon, and let $\mathcal{I} = \{i, i+1, \dots, i+N_p\}$ represent the set of time indices within the current prediction window. In the $i$-th time slot, the optimal solution for the variables $(\mathbf{q}[i'], \mathbf{v}[i'], \mathbf{w}_k[i'], \mathbf{m}[i'], \mathbf{\Theta}[i'], \mathbf{\Psi}[i'])$ over the receding horizon $[i, i+N_p]$ can be obtained by solving the following MPC optimization problem:
\begin{subequations} \label{prob:P2}
\begin{align}
    (\text{P2}) \quad & \max_{ \substack{\mathbf{q}[i'], \mathbf{v}[i'], \mathbf{w}_k[i'], \\ \mathbf{m}[i'], \mathbf{\Theta}[i'], \mathbf{\Psi}[i']} } \quad \sum_{i'=i}^{i+N_p} \bigg( \chi_1 \sum_{k \in \mathcal{K}} D_k[i'] \nonumber \\
    &  \qquad - \chi_2 \sum \|\mathbf{q}[i'] - \mathbf{q}_B\|^2 - \chi_3 P\big(\mathbf{v}[i']\big) \bigg), \nonumber \\
    \textrm{s.t.} \quad & \text{SNR}^r[i'] \geq \gamma^r, \label{eq:p2_snr} \\
    & \mathbf{q}[i'+1] = \mathbf{q}[i'] + \mathbf{v}[i'] t_c, \; i' \in \mathcal{I} \setminus \{i+N_p\}, \label{eq:p2_kinematics} \\
    & \left\| (v_x(i'), v_y(i')) \right\| \leq V_{\max}, \label{eq:p2_v_hor} \\
    & v_z(i') \leq U_{\max}, \label{eq:p2_v_ver} \\
    & \left\| \mathbf{v}[i'+1] - \mathbf{v}[i'] \right\| \leq a_{\max} t_c, \; i' \in \mathcal{I} \setminus \{i+N_p\}, \label{eq:p2_acc} \\
    & Z_{\min} \leq q_z(i') \leq Z_{\max}, \label{eq:p2_height} \\
    & \sum_{k=1}^{K} \|\mathbf{w}_k[i']\|^2 + \|\mathbf{m}[i']\|^2 \leq P_{\max}, \label{eq:p2_power} \\
    & \left| [\mathbf{\Theta}[i']]_{m,m} \right| = 1, \label{eq:p2_ris_theta} \\
    & \left| [\mathbf{\Psi}[i']]_{m,m} \right| = 1. \label{eq:p2_ris_psi}
\end{align}
\end{subequations}
Consequently, the infinite-horizon problem (P1) is recast into a finite-horizon MPC formulation (P2). Within this overarching MPC framework, solving (P2) at each receding step remains computationally intractable due to strongly coupled variables and non-convex constraints. To execute the MPC efficiently, we embed an AO algorithm that decouples (P2) into three tractable sub-problems, utilizing SCA to handle local non-convexities.
\begin{figure*}[!t]
\normalsize
\setcounter{MYtempeqncnt}{\value{equation}}
\setcounter{equation}{30} 

\begin{equation} \label{eq:rate_expansion}
\begin{split} 
    R_k[i'] = & \underbrace{ \log_2 \left( \sum_{c=1}^K \left| \mathbf{h}_{\mathrm{r}k}^H[i'] \mathbf{\Theta}[i'] \mathbf{H}_{\mathrm{dr}}^H[i'] \mathbf{w}_c[i'] \right|^2 + \left| \mathbf{h}_{\mathrm{r}k}^H[i'] \mathbf{\Theta}[i'] \mathbf{H}_{\mathrm{dr}}^H[i'] \mathbf{m}[i'] \right|^2 + \sigma_k^2[i'] \right) }_{R_k^1[i']} \\
    & - \underbrace{ \log_2 \left( \sum_{c=1, c \neq k}^K \left| \mathbf{h}_{\mathrm{r}k}^H[i'] \mathbf{\Theta}[i'] \mathbf{H}_{\mathrm{dr}}^H[i'] \mathbf{w}_c[i'] \right|^2 + \left| \mathbf{h}_{\mathrm{r}k}^H[i'] \mathbf{\Theta}[i'] \mathbf{H}_{\mathrm{dr}}^H[i'] \mathbf{m}[i'] \right|^2 + \sigma_k^2[i'] \right) }_{R_k^2[i']}.
\end{split}
\end{equation}

\setcounter{equation}{\value{MYtempeqncnt}}
\hrulefill
\vspace*{4pt}
\end{figure*}
\section{MPC-Based Beamforming and Trajectory Optimization}
\subsection{Problem Transformation}
First, we employ the matrix lifting technique to reformulate (P2). Let $\mathbf{W}_k[i'] = \mathbf{w}_k[i']\mathbf{w}_k^H[i'] \in \mathbb{C}^{N \times N}$ and $\mathbf{M}[i'] = \mathbf{m}[i']\mathbf{m}^H[i'] \in \mathbb{C}^{N \times N}$. Consequently, the transformed optimization problem, denoted as (P3), can be expressed as:
\allowdisplaybreaks
\begin{subequations} \label{prob:P3}
\begin{align}
    (\text{P3}) \: & \max_{ \substack{\mathbf{q}[i'], \mathbf{v}[i'], \mathbf{W}_k[i'], \\ \mathbf{M}[i'], \mathbf{\Theta}[i'], \mathbf{\Psi}[i']} } \: \sum_{i'=i}^{i+N_p} \bigg( \chi_1 \sum_{k \in \mathcal{K}} D_k[i'] \nonumber \\
    & \qquad \qquad - \chi_2 \sum \|\mathbf{q}[i'] - \mathbf{q}_B\|^2 - \chi_3 P\big(\mathbf{v}[i']\big) \bigg), \nonumber \\
    \textrm{s.t.} \quad & {\small\eqref{eq:p2_snr}, \eqref{eq:p2_kinematics}, \eqref{eq:p2_v_hor}, \eqref{eq:p2_v_ver}, \eqref{eq:p2_acc}, \eqref{eq:p2_height}, \eqref{eq:p2_ris_theta}, \eqref{eq:p2_ris_psi}}, \label{eq:p3_ref_p2} \\
    & \sum_{k=1}^{K} \mathrm{Tr}\big(\mathbf{W}_k[i']\big) + \mathrm{Tr}\big(\mathbf{M}[i']\big) \leq P_{\max}, \label{eq:p3_power} \\
    & \mathbf{W}_k[i'] \succeq 0, \quad \mathbf{M}[i'] \succeq 0, \label{eq:p3_sdp} \\
    & \mathrm{rank}\big(\mathbf{W}_k[i']\big) = 1, \quad \mathrm{rank}\big(\mathbf{M}[i']\big) = 1. \label{eq:p3_rank}
\end{align}
\end{subequations}
Constraints \eqref{eq:p3_sdp} and \eqref{eq:p3_rank} introduce semi-definite and rank-1 constraints, by applying matrix lifting to the transmit beamforming and artificial noise vectors. Nevertheless, problem (P3) remains highly non-convex and intractable to solve directly. This intractability stems from the strong non-linear coupling among the communication, sensing, and energy objectives, further compounded by the non-convex unit modulus constraints and coupled channel responses. To address this challenge, we decouple the original problem into three tractable sub-problems and solve them iteratively via an AO approach, as detailed in the subsequent sections.
\subsection{Proposed Solution}
In this paper, the original problem is decomposed into three sub-problems: UAV trajectory planning, beamforming and AN optimization, and RIS phase shift matrix design. These sub-problems are solved iteratively using an AO method. Furthermore, to address the uncertainty of channel state information and the dynamic environmental disturbances affecting the UAV, a 
MPC mechanism is introduced. By dynamically correcting the optimization strategy within a receding horizon, this mechanism effectively suppresses cumulative errors and enhances system robustness.
\subsubsection{UAV Trajectory Planning}

In the $i$-th time slot, given the fixed RIS uplink echo reflection coefficients $\mathbf{\Psi}[i]$, the downlink communication reflection coefficients $\mathbf{\Theta}[i]$, the beamforming vectors $\mathbf{w}_k[i]$, and the AN vector $\mathbf{m}[i]$, we optimize the UAV's flight trajectory. Our objective is to minimize the flight path length and power consumption while strictly satisfying secure communication and sensing constraints. Consequently, the trajectory optimization sub-problem (P4) is formulated as follows:
\allowdisplaybreaks
\begin{subequations} \label{prob:P4}
\begin{align}
    (\text{P4}) \: & \max_{ \mathbf{q}[i'], \mathbf{v}[i'] } \quad \sum_{i'=i}^{i+N_p} \bigg( \chi_1 \sum_{k \in \mathcal{K}} D_k[i'] - \chi_2 \sum \|\mathbf{q}[i'] - \mathbf{q}_B\|^2 \nonumber \\
    & \qquad \qquad - \chi_3 P\big(\mathbf{v}[i']\big) \bigg), \nonumber \\
    \textrm{s.t.} \quad & \eqref{eq:p2_snr}, \eqref{eq:p2_kinematics}, \eqref{eq:p2_v_hor}, \eqref{eq:p2_v_ver}, \eqref{eq:p2_acc}, \eqref{eq:p2_height}.
\end{align}
\end{subequations}
In (P4), both the propulsion power consumption term $P(\mathbf{v}[i'])$ and the achievable secrecy rate term $D_k[i']$ exhibit non-convex characteristics. To address this, we employ the following techniques to resolve the non-convexity.

First, based on \eqref{eq:power_model}, the power consumption of the UAV in the $i'$-th time slot can be expressed as:
\begin{equation} \label{eq:power_original}
\begin{aligned}
    P\big(\mathbf{v}[i']\big) = & \frac{W^2}{\sqrt{2}\rho S} \left( \|\mathbf{v}_h[i']\|^2 + \sqrt{\|\mathbf{v}_h[i']\|^4 + 4V_h^4} \right)^{\frac{1}{2}} \\
    & + W v_z(i') + \frac{\zeta \rho S}{8} \|\mathbf{v}_h[i']\|^3,
\end{aligned}
\end{equation}
where $W$ is the aircraft weight, $\rho$ is the air density, and $S$ is the rotor disc area. 
It can be observed from this power model that it exhibits non-convex characteristics due to the inclusion of high-order nonlinear terms, posing significant challenges in both theory and computation for directly solving the optimization problem. To overcome this difficulty, we perform a convex approximation to obtain its convex upper bound as follows\cite{T34}:
\begin{equation} \label{eq:power_upper_bound}
    P^{\text{ub}}\big(\mathbf{v}[i']\big) = \frac{W^2}{\sqrt{2}\rho S} \left( f_{i'}^n \big(\mathbf{v}_h[i']\big) \right)^{\frac{1}{2}} + W v_z(i') + \frac{\zeta \rho S}{8} \|\mathbf{v}_h[i']\|^3,
\end{equation}
where
\begin{subequations}
\begin{align}
    f_{i'}^n\big(\mathbf{v}_h[i']\big) = & f_{i'}^{-n}\big(\mathbf{v}_h[i']\big) + 2 \left\langle \mathbf{v}_h[i'], \mathbf{v}_h^n[i'] \right\rangle - \|\mathbf{v}_h^n[i']\|^2, \label{eq:f_n} \\
    f_{i'}^{-n}\big(\mathbf{v}_h[i']\big) = & \left( \|\mathbf{v}_h^n[i']\|^4 + 4V_h^4 \right)^{-\frac{1}{2}} \bigg( 4V_h^4 + \|\mathbf{v}_h^n[i']\|^2 \nonumber \\
    & \times \Big( 2 \langle \mathbf{v}_h[i'], \mathbf{v}_h^n[i'] \rangle - \|\mathbf{v}_h^n[i']\|^2 \Big) \bigg), \label{eq:f_minus_n}
\end{align}
\end{subequations}
where $\mathbf{v}_h^n[i']$ denotes the horizontal velocity of the UAV in the $n$-th iteration.Meanwhile, the achievable rate of the $k$-th legitimate user in the $i'$-th time slot can be expressed as \eqref{eq:rate_expansion}.

It is evident that $R_k[i']$ can be decomposed into the difference of two terms, $R_k[i'] = R_k^1[i'] - R_k^2[i']$, both of which are concave functions. To transform the objective into a concave function, we proceed with the following reformulation. Let $A[i']$ denote the interference-plus-noise term, defined as:
\addtocounter{equation}{1}
\begin{equation} \label{eq:your_equation_32}
\begin{split}
    A[i'] = & \sum_{c=1, c \neq k}^{K} \left| \mathbf{h}_{\mathrm{r}k}^H[i'] \mathbf{\Theta}[i'] \mathbf{H}_{\mathrm{dr}}^H[i'] \mathbf{w}_c[i'] \right|^2 \\
    & + \left| \mathbf{h}_{\mathrm{r}k}^H[i'] \mathbf{\Theta}[i'] \mathbf{H}_{\mathrm{dr}}^H[i'] \mathbf{m}[i'] \right|^2 + \sigma_k^2[i'].
\end{split}
\end{equation}
Since the channel gain within $A[i']$ varies as a function of the UAV's position, we apply the chain rule to differentiate $A[i']$, thereby deriving the gradient expression of $R_k^2\big(\mathbf{q}[i']\big)$ with respect to the position:
\begin{equation} \label{eq:gradient_R2}
    \nabla_{\mathbf{q}[i']} R_k^2\big(\mathbf{q}[i']\big) = \frac{1}{\ln 2 \cdot A[i']} \nabla_{\mathbf{q}[i']} A\big(\mathbf{q}[i']\big).
\end{equation}

At the $n$-th iteration point $\mathbf{q}^n[i']$, based on the first-order Taylor expansion, we construct the linear approximation $l_{R_k^2[i]}(\mathbf{q}[i])$ of $R_k^2[i]$ as follows:
\begin{equation} \label{eq:taylor_approx}
    l_{R_k^2}\big(\mathbf{q}[i']\big) = R_k^2\big(\mathbf{q}^n[i']\big) + \nabla_{\mathbf{q}^n[i']} R_k^2\big(\mathbf{q}^n[i']\big)^T \big(\mathbf{q}[i'] - \mathbf{q}^n[i']\big).
\end{equation}
Then, we define:
\begin{equation} \label{eq:R_new}
    R_k^{\text{new}} = R_k^1 - l_{R_k^2}\big(\mathbf{q}[i']\big).
\end{equation}
Through this method, $R_k[i']$ is transformed into a tractable form suitable for convex optimization processing. In the iterative optimization process, continuously updating $\mathbf{q}^n[i']$, we progressively solve the original problem.
\begin{figure*}[!t]
\normalsize
\begin{equation} \label{eq:eaves_rate_expansion}
    C_k[i'] = \underbrace{ \log_2 \Big( \big| \mathbf{h}_{\mathrm{re}}^H[i'] \mathbf{\Theta}[i'] \mathbf{H}_{\mathrm{dr}}^H[i'] \mathbf{w}_k[i'] \big|^2 + \big| \mathbf{h}_{\mathrm{re}}^H[i'] \mathbf{\Theta}[i'] \mathbf{H}_{\mathrm{dr}}^H[i'] \mathbf{m}[i'] \big|^2 + \sigma_{\mathrm{e}}^2[i'] \Big) }_{C_k^1[i']} 
    - \underbrace{ \log_2 \Big( \big| \mathbf{h}_{\mathrm{re}}^H[i'] \mathbf{\Theta}[i'] \mathbf{H}_{\mathrm{dr}}^H[i'] \mathbf{m}[i'] \big|^2 + \sigma_{\mathrm{e}}^2[i'] \Big) }_{C_k^2[i']}.
\end{equation}
\hrulefill
\vspace*{4pt}
\end{figure*}


Similarly, the eavesdropping rate $C_k[i']$ can be processed using  \eqref{eq:eaves_rate_expansion}-\eqref{eq:Ck_new} as follows. The gradient of $C_k^1$ with respect to the trajectory is given by:
\begin{equation} \label{eq:gradient_C1}
    \nabla_{\mathbf{q}[i']} C_k^1\big(\mathbf{q}[i']\big) = \frac{1}{\ln 2 \cdot B[i']} \nabla_{\mathbf{q}[i']} B\big(\mathbf{q}[i']\big),
\end{equation}
where $B[i']$ is defined as:
\begin{equation}
\begin{split}
    B[i'] = & \left| \mathbf{h}_{\mathrm{re}}^H[i'] \mathbf{\Theta}[i'] \mathbf{H}_{\mathrm{dr}}^H[i'] \mathbf{w}_k[i'] \right|^2 \\
    & + \left| \mathbf{h}_{\mathrm{re}}^H[i'] \mathbf{\Theta}[i'] \mathbf{H}_{\mathrm{dr}}^H[i'] \mathbf{m}[i'] \right|^2 + \sigma_{\mathrm{e}}^2[i'].
\end{split}
\end{equation}
Analogously, applying the first-order Taylor expansion at the $n$-th iteration point $\mathbf{q}^n[i']$, we construct the linear approximation $l_{C_k^1}(\mathbf{q}[i'])$ for $C_k^1[i']$ as:
\begin{equation} \label{eq:taylor_approx_C1}
    l_{C_k^1}\big(\mathbf{q}[i']\big) = C_k^1\big(\mathbf{q}^n[i']\big) + \nabla_{\mathbf{q}^n[i']} C_k^1\big(\mathbf{q}^n[i']\big)^T \big(\mathbf{q}[i'] - \mathbf{q}^n[i']\big).
\end{equation}
We then define the convex approximation term:
\begin{equation} \label{eq:Ck_new}
    C_k^{\text{new}}[i'] = l_{C_k^1}\big(\mathbf{q}[i']\big) - C_k^2[i'].
\end{equation}
Finally, we process $D_k[i']$ as follows to transform it into a concave function:
\begin{equation} \label{eq:Dk_new}
    D_k^{\text{new}}[i'] = R_k^{\text{new}}[i'] - C_k^{\text{new}}[i'].
\end{equation}
Following the aforementioned approximation procedures, during the $n$-th iteration, the original non-convex (P4) is reformulated into the following approximate convex optimization problem, denoted as (P4.1):
\allowdisplaybreaks
\begin{subequations} \label{prob:P4_1}
\small
\begin{align}
    (\text{P4.1}) \: & \max_{ \mathbf{q}[i'], \mathbf{v}[i'] } \quad \sum_{i'=i}^{i+N_p} \bigg( \chi_1 \sum_{k \in \mathcal{K}} D_k^{\text{new}}[i'] - \chi_2 \sum \|\mathbf{q}[i'] - \mathbf{q}_B\|^2 \nonumber \\
    & \qquad \qquad - \chi_3 P^{\text{ub}}\big(\mathbf{v}[i']\big) \bigg), \nonumber \\
    \textrm{s.t.} \quad & \eqref{eq:p2_snr}, \eqref{eq:p2_kinematics}, \eqref{eq:p2_v_hor}, \eqref{eq:p2_v_ver}, \eqref{eq:p2_acc}, \eqref{eq:p2_height}.
\end{align}
\end{subequations}
Thus, the objective function comprises concave functions and negative convex functions, rendering it overall concave with respect to the optimization variables. This allows us to utilize standard convex solvers (e.g., CVX) for efficient solutions. By updating the Taylor expansion point based on the previous trajectory, we iteratively solve this problem until convergence.
\subsubsection{Beamforming and AN Optimization}
In this subsection, we focus on optimizing the design of the transmit beamforming vectors and the AN covariance matrix at the DFRC BS. Specifically, given the UAV's position, the RIS reflection coefficients for the uplink echo signal $\mathbf{\Psi}[i']$, and the reflection coefficients for the downlink sensing-communication signal $\mathbf{\Theta}[i']$, we proceed to optimize the DFRC BS transmission. Consequently, the optimization problem regarding the transmit beamforming and AN design can be formulated as (P5):
\begin{subequations}
\begin{align}
(\text{P5}): \quad & \max_{\{\mathbf{W}_k[i'], \mathbf{M}[i']\}} \sum_{i'=i}^{i+N_p} \chi_1 \sum_{k \in \mathcal{K}} D_k[i'] \nonumber \\
\text{s.t.} \quad & \text{SNR}^r[i'] \geq \gamma^r, \label{prob:P5_c1} \\
& \sum_{k=1}^{K} \text{Tr}\left(\mathbf{W}_k[i']\right) + \text{Tr}\left(\mathbf{M}[i']\right) \leq P_{\max}, \label{prob:P5_c2} \\
& \mathbf{W}_k[i'] \succeq \mathbf{0}, \quad \mathbf{M}[i'] \succeq \mathbf{0}, \label{prob:P5_c3} \\
& \text{rank}\left(\mathbf{W}_k[i']\right) = 1, \quad \text{rank}\left(\mathbf{M}[i']\right) = 1, \label{prob:P5_c4}
\end{align}
\end{subequations}
\begin{figure*}[!t]
\normalsize
\begin{align}
    R_k[i'] &= \log_2 \Bigg( \frac{\sum_{c=1}^{K} \mathrm{Tr}\big(\mathbf{H}_k[i'] \mathbf{W}_c[i']\big) + \mathrm{Tr}\big(\mathbf{H}_k[i'] \mathbf{M}[i']\big) + \sigma_k^2}{\sum_{c \neq k}^{K} \mathrm{Tr}\big(\mathbf{H}_k[i'] \mathbf{W}_c[i']\big) + \mathrm{Tr}\big(\mathbf{H}_k[i'] \mathbf{M}[i']\big) + \sigma_k^2} \Bigg) \nonumber \\
    &= \underbrace{\log_2 \bigg( \sum_{c=1}^{K} \mathrm{Tr}\big(\mathbf{H}_k[i'] \mathbf{W}_c[i']\big) + \mathrm{Tr}\big(\mathbf{H}_k[i'] \mathbf{M}[i']\big) + \sigma_k^2 \bigg)}_{R_k^3[i']} 
    - \underbrace{\log_2 \bigg( \sum_{c \neq k}^{K} \mathrm{Tr}\big(\mathbf{H}_k[i'] \mathbf{W}_c[i']\big) + \mathrm{Tr}\big(\mathbf{H}_k[i'] \mathbf{M}[i']\big) + \sigma_k^2 \bigg)}_{R_k^4[i']}, \label{eq:rate_legitimate_diff} \\
    C_k[i'] &= \log_2 \Bigg( 1 + \frac{\mathrm{Tr}\big(\mathbf{H}_{\mathrm{e}}[i'] \mathbf{W}_k[i']\big)}{\mathrm{Tr}\big(\mathbf{H}_{\mathrm{e}}[i'] \mathbf{M}[i']\big) + \sigma_{\mathrm{e}}^2} \Bigg) \!=\! \underbrace{\log_2 \Big( \mathrm{Tr}\big(\mathbf{H}_{\mathrm{e}}[i'] \mathbf{M}[i']\big) \!+\! \sigma_{\mathrm{e}}^2 \!+\! \mathrm{Tr}\big(\mathbf{H}_{\mathrm{e}}[i'] \mathbf{W}_k[i']\big) \Big)}_{C_k^3[i']} \!-\! \underbrace{\log_2 \Big( \mathrm{Tr}\big(\mathbf{H}_{\mathrm{e}}[i'] \mathbf{M}[i']\big) \!+\! \sigma_{\mathrm{e}}^2 \Big)}_{C_k^4[i']}, \label{eq:rate_eavesdropper_diff}
\end{align}
\hrulefill
\vspace*{4pt}
\end{figure*}
Given the inherent non-convexity of the problem and the rank-1 constraints, we adopt the semidefinite relaxation (SDR) method to address it. Specifically, we first reformulate the achievable rate of the legitimate user and the interception rate of the potential eavesdropper as follows:
{\small
\begin{align}
    R_k[i'] &= \log_2 \left( 1 + \frac{\left| \mathbf{h}_k[i'] \mathbf{w}_k[i'] \right|^2}{\sum\limits_{c \neq k}^K \left| \mathbf{h}_k[i'] \mathbf{w}_c[i'] \right|^2 + \left| \mathbf{h}_k[i'] \mathbf{m}[i'] \right|^2 + \sigma_k^2} \right), 
\end{align}
\begin{align}
    C_k[i'] &= \log_2 \left( 1 + \frac{\left| \mathbf{h}_{\mathrm{e}}[i'] \mathbf{w}_k[i'] \right|^2}{\left| \mathbf{h}_{\mathrm{e}}[i'] \mathbf{m}[i'] \right|^2 + \sigma_{\mathrm{e}}^2} \right),
\end{align}
}
where $\mathbf{h}_k[i'] = \mathbf{h}_{\mathrm{r}k}^H[i'] \mathbf{\Theta}[i'] \mathbf{H}_{\mathrm{dr}}^H[i'] \in \mathbb{C}^{1 \times N}$ and $\mathbf{h}_{\mathrm{e}}[i'] = \mathbf{h}_{\mathrm{re}}^H[i'] \mathbf{\Theta}[i'] \mathbf{H}_{\mathrm{dr}}^H[i'] \in \mathbb{C}^{1 \times N}$. We further define $\mathbf{H}_k[i'] = \mathbf{h}_k^H[i']\mathbf{h}_k[i'] \in \mathbb{C}^{N \times N}$ and $\mathbf{H}_{\mathrm{e}}[i'] = \mathbf{h}_{\mathrm{e}}^H[i']\mathbf{h}_{\mathrm{e}}[i'] \in \mathbb{C}^{N \times N}$, which allows us to transform the legitimate user's rate and the potential eavesdropper's interception rate into  \eqref{eq:rate_legitimate_diff} and \eqref{eq:rate_eavesdropper_diff}.
It can be observed that $R_k^3[i']$, $R_k^4[i']$, $C_k^3[i']$, and $C_k^4[i']$ are all concave functions with respect to the variables $\mathbf{W}_k[i']$ and $\mathbf{M}[i']$. Based on this observation, we proceed to transform the objective function as described below.
The gradients of the aforementioned terms with respect to the variables $\mathbf{M}[i']$ and $\mathbf{W}_k[i']$ are derived as follows:
\begin{subequations} \label{eq:all_gradients}
\allowdisplaybreaks
\begin{align}
    \nabla_{\mathbf{M}[i']} R_k^4 [\mathbf{M}[i']] &= \frac{\mathbf{H}_k[i']}{\ln 2 \cdot E_k[i']}, \label{eq:grad_R4_M} \\
    \nabla_{\mathbf{W}_k[i']} R_k^4 [\mathbf{W}_k[i']] &= \mathbf{0}, \label{eq:grad_R4_W} \\
    \nabla_{\mathbf{M}[i']} C_k^3 [\mathbf{M}[i']] &= \frac{\mathbf{H}_{\mathrm{e}}[i']}{\ln 2 \cdot F_k[i']}, \label{eq:grad_C3_M} \\
    \nabla_{\mathbf{W}_k[i']} C_k^3 [\mathbf{W}_k[i']] &= \frac{\mathbf{H}_{\mathrm{e}}[i']}{\ln 2 \cdot F_k[i']}. \label{eq:grad_C3_W}
\end{align}
\end{subequations}
where,we define the denominator terms $E_k[i']$ and $F_k[i']$ as:
\begin{subequations} \label{eq:denominator_terms}
\begin{align}
    E_k[i'] &\!=\! \sum_{c \neq k}^{K} \mathrm{Tr}\big(\mathbf{H}_k[i'] \mathbf{W}_c[i']\big)\! +\! \mathrm{Tr}\big(\mathbf{H}_k[i'] \mathbf{M}[i']\big) \!+\! \sigma_k^2, \label{eq:E_k} \\
    F_k[i'] &= \mathrm{Tr}\big(\mathbf{H}_{\mathrm{e}}[i'] \mathbf{M}[i']\big) + \sigma_{\mathrm{e}}^2 + \mathrm{Tr}\big(\mathbf{H}_{\mathrm{e}}[i'] \mathbf{W}_k[i']\big). \label{eq:F_k}
\end{align}
\end{subequations}

We utilize \eqref{eq:grad_R4_M} and \eqref{eq:grad_R4_W} to process the second term of \eqref{eq:rate_legitimate_diff}. Specifically, let $\mathbf{W}_k^n[i']$ and $\mathbf{M}^n[i']$ denote the values at the $n$-th iteration. By performing the expansion around the point $(\mathbf{W}_k^n[i'], \mathbf{M}^n[i'])$, we obtain:
\begin{align}
    &R_k^{4,l}\big[\mathbf{W}_k^n[i'], \mathbf{M}^n[i']\big] = R_k^4\big[\mathbf{W}_k^n[i'], \mathbf{M}^n[i']\big] \nonumber \\
    &\quad + \nabla_{\mathbf{M}^n[i']} R_k^4 \big[\mathbf{W}_k^n[i'], \mathbf{M}^n[i']\big] \cdot \big(\mathbf{M}[i'] - \mathbf{M}^n[i']\big), 
\end{align}
\begin{equation} \label{eq:R_LB}
    R_k^{\mathbf{M}[i'], \mathbf{W}_k[i']}[i'] = R_k^3 \big[\mathbf{M}[i'], \mathbf{W}_k[i']\big] - R_k^{4,l} \big[\mathbf{M}[i'], \mathbf{W}_k[i']\big].
\end{equation}
Using \eqref{eq:grad_C3_M} and \eqref{eq:grad_C3_W}, we process the first term of \eqref{eq:rate_eavesdropper_diff} as:
\begin{align}
    & C_k^{3,l}\big[\mathbf{W}_k[i'], \mathbf{M}[i']\big] = C_k^3\big[\mathbf{W}_k^n[i'], \mathbf{M}^n[i']\big] \nonumber \\
    & + \nabla_{\mathbf{W}_k^n[i']} C_k^3 \big[\mathbf{W}_k^n[i'], \mathbf{M}^n[i']\big] \cdot \big(\mathbf{W}_k[i'] - \mathbf{W}_k^n[i']\big) \label{eq:taylor_C3} \\
    & + \nabla_{\mathbf{M}^n[i']} C_k^3 \big[\mathbf{W}_k^n[i'], \mathbf{M}^n[i']\big] \cdot \big(\mathbf{M}[i'] - \mathbf{M}^n[i']\big), \nonumber
\end{align}
\begin{equation} \label{eq:C_surrogate}
    C_k^{\mathbf{W}_k[i'], \mathbf{M}[i']}[i'] = C_k^{3,l}\big[\mathbf{W}_k[i'], \mathbf{M}[i']\big] - C_k^4\big[\mathbf{W}_k[i'], \mathbf{M}[i']\big].
\end{equation}
Then, we have transformed the objective function into a concave function as follows:
\begin{equation} \label{eq:D_k_final}
    D_k[i'] = R_k^{\mathbf{W}_k[i'], \mathbf{M}[i']}[i'] - C_k^{\mathbf{W}_k[i'], \mathbf{M}[i']}[i'].
\end{equation}
Furthermore, for the positive semidefinite matrices $\mathbf{W}_k[i']$ and $\mathbf{M}[i']$, the rank-1 constraints $\text{rank}(\mathbf{W}_k[i']) = 1$ and $\text{rank}(\mathbf{M}[i']) = 1$ in \eqref{prob:P5_c4} are equivalent to the difference of two convex functions, expressed as:
\begin{subequations} \label{eq:rank_equivalence}
\begin{align}
    \text{rank}(\mathbf{W}_k[i']) = 1 &\Leftrightarrow \text{Tr}(\mathbf{W}_k[i']) - \|\mathbf{W}_k[i']\|_2 = 0, \label{eq:rank_W_equiv} \\
    \text{rank}(\mathbf{M}[i']) = 1 &\Leftrightarrow \text{Tr}(\mathbf{M}[i']) - \|\mathbf{M}[i']\|_2 = 0, \label{eq:rank_M_equiv}
\end{align}
\end{subequations}
where $\|\mathbf{W}_k[i']\|_2 = \sigma_1(\mathbf{W}_k[i'])$ and $\|\mathbf{M}[i']\|_2 = \sigma_1(\mathbf{M}[i'])$ denote the spectral norms, with $\sigma_1(\cdot)$ representing the largest singular value of the matrix. Therefore, we incorporate these functions as penalty terms into the objective function of (P5), formulating the following optimization problem (P5.1):
\begin{subequations}
\begin{align}
    & (\text{P5.1}) \max_{\mathbf{W}_k[i'], \mathbf{M}[i']} \sum_{i'=i}^{i+N_p} \chi_1 \sum_{k \in \mathcal{K}} D_k^{\mathbf{W}_k[i'], \mathbf{M}[i']}[i'] \nonumber \\
    & \qquad \qquad - \tau_1 \Big( \text{Tr}(\mathbf{W}_k[i']) - \|\mathbf{W}_k[i']\|_2 \Big) \nonumber \\
    & \qquad \qquad - \tau_2 \Big( \text{Tr}(\mathbf{M}[i']) - \|\mathbf{M}[i']\|_2 \Big) \nonumber \\
    & \text{s.t.} \qquad \eqref{prob:P5_c1}, \eqref{prob:P5_c2}, \eqref{prob:P5_c3}, \label{eq:P5_1_constraints}
\end{align}
\end{subequations}
where $\tau_1$ and $\tau_2$ are the penalty factors associated with the rank-1 constraints. Since the spectral norms $\|\mathbf{W}_k[i']\|_2$ and $\|\mathbf{M}[i']\|_2$ are convex,(P5.1) remains non-convex overall. To address this, we employ the DC programming approach. Let $n$ denote the current iteration index. We proceed to solve the (P5.2):
\begin{subequations}
\begin{align}
    & (\text{P5.2}) \max_{\mathbf{W}_k[i'], \mathbf{M}[i']} \sum_{i'=i}^{i+N_p} \chi_1 \sum_{k \in \mathcal{K}} D_k^{\mathbf{W}_k[i'], \mathbf{M}[i']}[i'] \nonumber \\
    & \qquad \qquad - \tau_1 \Big( \text{Tr}(\mathbf{W}_k[i']) - \partial \big\langle \|\mathbf{W}_k[i']\|_2^{n-1}, \mathbf{W}_k[i'] \big\rangle \Big) \nonumber \\
    & \qquad \qquad - \tau_2 \Big( \text{Tr}(\mathbf{M}[i']) - \partial \big\langle \|\mathbf{M}[i']\|_2^{n-1}, \mathbf{M}[i'] \big\rangle \Big) \nonumber \\
    & \text{s.t.} \qquad \eqref{prob:P5_c1}, \eqref{prob:P5_c2}, \eqref{prob:P5_c3}, \label{eq:P5_2_constraints}
\end{align}
\end{subequations}
where $\|\mathbf{W}_k[i']\|_2^{n-1}$ and $\|\mathbf{M}[i']\|_2^{n-1}$ are the solutions from the $(n-1)$-th iteration, and $\partial \big\langle \|\mathbf{W}_k[i']\|_2^{n-1} \big\rangle$ and $\partial \big\langle \|\mathbf{M}[i']\|_2^{n-1} \big\rangle$ denote the subgradients of the spectral norms at the $(n-1)$-th iteration. (P5.2) is a semidefinite program (SDP), which can be efficiently solved via CVX.
\subsubsection{RIS Phase Shift Matrix Design}
In this subsection, we aim to separately optimize the downlink phase shift matrix $\mathbf{\Theta}[i']$ and the uplink echo signal phase shift matrix $\mathbf{\Psi}[i']$ for the aerial RIS. First, regarding the design of the downlink phase shift matrix $\mathbf{\Theta}[i']$, given the aerial RIS's three-dimensional trajectory, the DFRC BS beamforming vectors $\{\mathbf{w}_k[i']\}$ and $\mathbf{m}[i']$, and the uplink phase shift matrix $\mathbf{\Psi}[i']$, the optimization problem can be reformulated as (P6):
\begin{subequations}
\begin{align}
    & (\text{P6}) \quad \max_{\mathbf{\Theta}[i']} \sum_{i'=i}^{i+N_p} \chi_1 \sum_{k \in \mathcal{K}} D_k[i'] \nonumber \\
    & \text{s.t.} \qquad \text{SNR}^r[i'] \geq \gamma^r, \label{eq:P6_snr} \\
    & \qquad \quad \left| [\mathbf{\Theta}[i']]_{m,m} \right| = 1, \label{eq:P6_modulus}
\end{align}
\end{subequations}
According to \eqref{eq:theta}, let $\mathbf{g}[i'] = \big[e^{j\alpha_1[i']}, \dots, e^{j\alpha_M[i']}\big]^T$. We define $\mathbf{G}[i'] = \mathbf{g}[i']\mathbf{g}^H[i'] \in \mathbb{C}^{M \times M}$, which satisfies $\text{rank}(\mathbf{G}[i']) = 1$. Consequently, the optimization problem can be transformed into (P6.1), formulated as follows:
\begin{subequations}
\begin{align}
    & (\text{P6.1}) \quad \max_{\mathbf{\Theta}[i']} \sum_{i'=i}^{i+N_p} \chi_1 \sum_{k \in \mathcal{K}} D_k[i'] \nonumber \\
    & \text{s.t.} \qquad \text{SNR}^r[i'] \geq \gamma^r, \label{eq:P6_1_snr} \\
    & \qquad \quad \mathbf{G}_{m,m}[i'] = 1, \label{eq:P6_1_diagonal} \\
    & \qquad \quad \mathbf{G}[i'] \succeq \mathbf{0}, \label{eq:P6_1_sdp} \\
    & \qquad \quad \text{rank}\big(\mathbf{G}[i']\big) = 1. \label{eq:P6_1_rank}
\end{align}
\end{subequations}
\begin{figure*}[!t]
\normalsize
\begin{equation} \label{eq:rate_legitimate_G}
    R_k[i'] = \underbrace{\log_2 \bigg( \sum_{c=1}^{K} \mathrm{Tr}\big(\mathbf{R}_{k\mathrm{d}}[i']\mathbf{G}[i']\big) + \mathrm{Tr}\big(\mathbf{R}_{k\mathrm{m}}[i']\mathbf{G}[i']\big) + \sigma_k^2 \bigg)}_{R_k^5[i']} 
    - \underbrace{\log_2 \bigg( \sum_{c \neq k}^{K} \mathrm{Tr}\big(\mathbf{R}_{k\mathrm{d}}[i']\mathbf{G}[i']\big) + \mathrm{Tr}\big(\mathbf{R}_{k\mathrm{m}}[i']\mathbf{G}[i']\big) + \sigma_k^2 \bigg)}_{R_k^6[i']},
\end{equation}
\hrulefill 
\vspace*{4pt}
\end{figure*}
We reformulate the legitimate user's communication rate $R_k[i']$, the potential eavesdropper's interception rate, and the sensing performance metric as \eqref{eq:rate_legitimate_G} and \eqref{eq:rate_eavesdropper_G}:
\allowdisplaybreaks
\begin{align}
    C_k[i'] &= \underbrace{\log_2 \left( \text{Tr}\big(\mathbf{R}_{\mathrm{em}}[i']\mathbf{G}[i']\big) + \sigma_{\mathrm{e}}^2 + \text{Tr}\big(\mathbf{R}_{\mathrm{e}k}[i']\mathbf{G}[i']\big) \right)}_{C_k^5[i']} \nonumber \\
    &\quad - \underbrace{\log_2 \left( \text{Tr}\big(\mathbf{R}_{\mathrm{em}}[i']\mathbf{G}[i']\big) + \sigma_{\mathrm{e}}^2 \right)}_{C_k^6[i']}, \label{eq:rate_eavesdropper_G} \\
\text{SNR}^{\mathrm{r}}[i'] &= \frac{\sum\limits_{k=1}^{K} \text{Tr}\big(\mathbf{R}_{\mathrm{r}k}[i']\mathbf{G}[i']\big) + \text{Tr}\big(\mathbf{R}_{\mathrm{rm}}[i']\mathbf{G}[i']\big)}{\sigma_{\mathrm{r}}^2[i]}, \label{eq:snr_radar_G}
\end{align}
where the auxiliary matrices are defined as:
\begin{subequations} \label{eq:all_R_matrices}
\begin{align}
    \mathbf{R}_{k\mathrm{d}}[i'] &= \big(\mathbf{h}_{\mathrm{r}k}[i']\mathbf{h}_{\mathrm{r}k}^H[i']\big) \odot \big(\mathbf{H}_{\mathrm{dr}}^H[i'] \mathbf{W}_k[i'] \mathbf{H}_{\mathrm{dr}}[i']\big)^T, \label{eq:def_R_kd} \\
    \mathbf{R}_{k\mathrm{m}}[i'] &= \big(\mathbf{h}_{\mathrm{r}k}[i']\mathbf{h}_{\mathrm{r}k}^H[i']\big) \odot \big(\mathbf{H}_{\mathrm{dr}}^H[i'] \mathbf{M}[i'] \mathbf{H}_{\mathrm{dr}}[i']\big)^T, \label{eq:def_R_km} \\
    \mathbf{R}_{\mathrm{e}k}[i'] &= \big(\mathbf{h}_{\mathrm{re}}[i']\mathbf{h}_{\mathrm{re}}^H[i']\big) \odot \big(\mathbf{H}_{\mathrm{dr}}^H[i'] \mathbf{W}_k[i'] \mathbf{H}_{\mathrm{dr}}[i']\big)^T, \label{eq:def_R_ek} \\
    \mathbf{R}_{\mathrm{em}}[i'] &= \big(\mathbf{h}_{\mathrm{re}}[i']\mathbf{h}_{\mathrm{re}}^H[i']\big) \odot \big(\mathbf{H}_{\mathrm{dr}}^H[i'] \mathbf{M}[i'] \mathbf{H}_{\mathrm{dr}}[i']\big)^T, \label{eq:def_R_em} \\
    \mathbf{R}_{\mathrm{r}k}[i'] &= \big(\mathbf{H}_{\mathrm{dr}}^H[i']\mathbf{H}_{\mathrm{dr}}[i']\big) \odot \big(\mathbf{h}_{\mathrm{re}}[i']\mathbf{h}_{\mathrm{re}}^H[i'] \mathbf{\Psi}[i'] \mathbf{H}_{\mathrm{dr}}^H[i'] \notag \\
    &\quad \times \mathbf{W}_k[i'] \mathbf{H}_{\mathrm{dr}}[i'] \mathbf{\Psi}^H[i'] \mathbf{h}_{\mathrm{re}}[i']\mathbf{h}_{\mathrm{re}}^H[i']\big)^T, \label{eq:def_R_rk} \\
    \mathbf{R}_{\mathrm{rm}}[i'] &= \big(\mathbf{H}_{\mathrm{dr}}^H[i']\mathbf{H}_{\mathrm{dr}}[i']\big) \odot \big(\mathbf{h}_{\mathrm{re}}[i']\mathbf{h}_{\mathrm{re}}^H[i'] \mathbf{\Psi}[i'] \mathbf{H}_{\mathrm{dr}}^H[i'] \notag \\
    &\quad \times \mathbf{M}[i'] \mathbf{H}_{\mathrm{dr}}[i'] \mathbf{\Psi}^H[i'] \mathbf{h}_{\mathrm{re}}[i']\mathbf{h}_{\mathrm{re}}^H[i']\big)^T. \label{eq:def_R_rm}
\end{align}
\end{subequations}
Both the objective function $D_k[i']$ and the rank-1 constraint \eqref{eq:P6_1_rank} induce non-convexity in the optimization problem, which we address sequentially. First, regarding the processing of the objective function, we employ a method similar to the previous section. It is evident that $R_k^6[i']$ and $C_k^5[i']$ are concave functions with respect to the variable $\mathbf{G}[i']$. Based on this, we derive the gradients of these terms as follows:
\begin{align}
&\nabla_{\mathbf{G}[i']} R_k^6 [\mathbf{G}[i']] = \frac{1}{\ln 2} \left( \sum\limits_{c \neq k}^{K} \mathbf{R}_{k\mathrm{d}}[i'] + \mathbf{R}_{k\mathrm{m}}[i'] \right) \notag \\
&\times \left( \sum\limits_{c \neq k}^{K} \mathrm{Tr}\big(\mathbf{R}_{k\mathrm{d}}[i']\mathbf{G}[i']\big) + \mathrm{Tr}\big(\mathbf{R}_{k\mathrm{m}}[i']\mathbf{G}[i']\big) + \sigma_k^2 \right)^{-1}, \label{eq:grad_R6_G} \\
&\nabla_{\mathbf{G}[i']} C_k^5 [\mathbf{G}[i']] = \frac{1}{\ln 2} \left( \mathbf{R}_{\mathrm{em}}[i'] + \mathbf{R}_{\mathrm{e}k}[i'] \right) \notag \\
&\times \left( \mathrm{Tr}\big(\mathbf{R}_{\mathrm{em}}[i']\mathbf{G}[i']\big) + \sigma_{\mathrm{e}}^2 + \mathrm{Tr}\big(\mathbf{R}_{\mathrm{e}k}[i']\mathbf{G}[i']\big) \right)^{-1}. \label{eq:grad_C5_G}
\end{align}
Using \eqref{eq:grad_R6_G} to process the second term of \eqref{eq:rate_legitimate_G}, let $\mathbf{G}^n[i']$ denote the value of $\mathbf{G}[i']$ at the $n$-th iteration. By performing the first-order Taylor expansion, the second term of \eqref{eq:rate_legitimate_G} can be approximated as:
\begin{align}
    R_k^{6,l}\big[\mathbf{G}[i']\big] &= R_k^6\big[\mathbf{G}^n[i']\big] \nonumber \\
    &\quad + \text{Tr}\Big( \big(\nabla_{\mathbf{G}^n[i']} R_k^6\big)^H \big(\mathbf{G}[i'] - \mathbf{G}^n[i']\big) \Big), \label{eq:taylor_R6}
\end{align}
\eqref{eq:rate_legitimate_G} can be transformed into a concave function:
\begin{equation} \label{eq:R_surrogate_G}
    R_k^{\mathbf{G}[i']}[i'] = R_k^5[i'] - R_k^{6,l}[i'],
\end{equation}
Similarly, we transform \eqref{eq:rate_eavesdropper_G} as follows:
\begin{align}
    C_k^{5,l}\big[\mathbf{G}[i']\big] &= C_k^5\big[\mathbf{G}^n[i']\big] \nonumber \\
    &\quad + \text{Tr}\Big( \big(\nabla_{\mathbf{G}^n[i']} C_k^5\big)^H \big(\mathbf{G}[i'] - \mathbf{G}^n[i']\big) \Big), \label{eq:taylor_C5} \\
    C_k^{\mathbf{G}[i']}[i'] &= C_k^{5,l}[i'] - C_k^6[i']. \label{eq:C_surrogate_G}
\end{align}
Thus, \eqref{eq:C_surrogate_G} becomes a concave function. So far, the objective function can be expressed as:
\begin{equation} \label{eq:D_surrogate_G}
    D_k[i'] = R_k^{\mathbf{G}[i']}[i'] - C_k^{\mathbf{G}[i']}[i'].
\end{equation}
We have transformed the objective function into a concave function, which allows for direct resolution. Next, we address the rank-1 constraint. A rank-1 implies that the matrix has only one non-zero singular value. The non-convexity of the rank function makes the optimization problem with this constraint challenging to solve. Therefore, to further resolve the non-convexity in problem (P6), we employ the aforementioned method to formulate the rank-1 constraint \eqref{eq:P6_1_rank} as the difference of two convex functions:
\begin{equation} \label{eq:rank_G_equiv}
    \text{rank}(\mathbf{G}[i']) = 1 \iff \text{Tr}(\mathbf{G}[i']) - \|\mathbf{G}[i']\|_2 = 0,
\end{equation}
where $\|\mathbf{G}[i']\|_2$ denotes the spectral norm. Consequently, we add this equation as a penalty term to problem (P6.1), transforming it into (P6.2):
\begin{subequations}
\begin{align}
    & (\text{P6.2}) \quad \max_{\mathbf{\Theta}[i']} \sum\limits_{i'=i}^{i+N_p} \chi_1 \sum\limits_{k \in \mathcal{K}} D_k[i'] - \zeta \Big( \text{Tr}(\mathbf{G}[i']) - \|\mathbf{G}[i']\|_2 \Big) \nonumber \\
    & \text{s.t.} \qquad \eqref{eq:P6_1_snr}, \eqref{eq:P6_1_diagonal}, \eqref{eq:P6_1_sdp}. \label{eq:P6_2_constraints}
\end{align}
\end{subequations}
where $\zeta$ is the penalty factor associated with the rank-1 constraint. Since $\|\mathbf{G}[i']\|_2$ is convex, (P6.2) remains non-convex overall. We continue to utilize the DC programming approach. Let $n$ denote the current iteration index, and we solve the following (P6.3):
\begin{subequations}
\begin{align}
    & (\text{P6.3}) \quad \max_{\mathbf{\Theta}[i']} \sum\limits_{i'=i}^{i+N_p} \chi_1 \sum\limits_{k \in \mathcal{K}} D_k[i'] \nonumber \\
    & \qquad \qquad - \zeta \Big( \text{Tr}(\mathbf{G}[i']) - \partial \big\langle \|\mathbf{G}[i']\|_2^{n-1}, \mathbf{G}[i'] \big\rangle \Big) \nonumber \\
    & \text{s.t.} \qquad \eqref{eq:P6_1_snr}, \eqref{eq:P6_1_diagonal}, \eqref{eq:P6_1_sdp}. \label{eq:P6_2_constraints}
\end{align}
\end{subequations}
where $\|\mathbf{G}[i']\|_2^{n-1}$ is the solution obtained from the $(n-1)$-th iteration, and $\partial \big\langle \|\mathbf{G}[i']\|_2^{n-1} \big\rangle$ denotes the subgradient of the spectral norm at the $(n-1)$-th iteration.Under the $n$-th iteration, the objective function is transformed into an SDP optimization problem, which can be efficiently solved using standard SDP solvers such as CVX.

Next, we need to optimize the uplink phase shift matrix $\mathbf{\Psi}[i']$. Similarly, assuming the UAV's three-dimensional trajectory, the DFRC BS beamforming vectors $\{\mathbf{w}_k[i']\}$, $\mathbf{m}[i']$, and the downlink phase shift matrix $\mathbf{\Theta}[i']$ are given, the optimization problem can be formulated as a feasibility check problem, expressed as (P7):
\begin{subequations}
\begin{align}
    & (\text{P7}) \quad \text{find} \quad \mathbf{\Psi}[i'] \nonumber \\
    & \text{s.t.} \qquad \left| [\mathbf{\Psi}[i']]_{m,m} \right| = 1. \label{eq:P7_modulus}
\end{align}
\end{subequations}
According to \eqref{eq:psi}, let $\mathbf{p}[i'] = \big[e^{j\beta_1[i']}, \dots, e^{j\beta_M[i']}\big]^T$. We then define $\mathbf{P}[i'] = \mathbf{p}[i']\mathbf{p}^H[i'] \in \mathbb{C}^{M \times M}$, which satisfies $\text{rank}(\mathbf{P}[i']) = 1$. Consequently, the optimization problem (P7) can be transformed into (P7.1):
\begin{subequations}
\begin{align}
    & (\text{P7.1}) \quad \text{find} \quad \mathbf{P}[i'] \nonumber \\
    & \text{s.t.} \qquad \mathbf{P}_{m,m}[i'] = 1, \label{eq:P7_1_diagonal} \\
    & \qquad \quad \text{rank}(\mathbf{P}[i']) = 1, \quad \mathbf{P}[i'] \succeq \mathbf{0}. \label{eq:P7_1_rank}
\end{align}
\end{subequations}
It should be noted that the first term in constraint \eqref{eq:P7_1_rank} is a rank-1 constraint, which is still a non-convex constraint. Similar to the previous solution, optimization problem (P7.1) can be reformulated as:
\begin{subequations}
\begin{align}
    & (\text{P7.2}) \quad \min \sum_{i'=i}^{i+N_p} \text{Tr}\big(\mathbf{P}[i']\big) - \|\mathbf{P}[i']\|_2 \nonumber \\
    & \text{s.t.} \qquad \mathbf{P}_{m,m}[i'] = 1, \label{eq:P7_2_diagonal} \\
    & \qquad \quad \mathbf{P}[i'] \succeq \mathbf{0}. \label{eq:P7_2_sdp}
\end{align}
\end{subequations}
The objective function of optimization problem (P7.2) is still non-convex. Similarly, employing the method used in (P3), we linearize the concave term $-\|\mathbf{P}[i']\|_2$. Therefore, at the $n$-th iteration, we need to solve the following problem:
\begin{subequations}
\begin{align}
    & (\text{P7.3}) \quad \min \sum_{i'=i}^{i+N_p} \text{Tr}\big(\mathbf{P}[i']\big) - \Big( \partial \big\langle \|\mathbf{P}[i']\|_2^{n-1}, \mathbf{P}[i'] \big\rangle \Big) \nonumber \\
    & \text{s.t.} \qquad \eqref{eq:P7_2_diagonal}, \eqref{eq:P7_2_sdp}. \label{eq:P7_3_constraints}
\end{align}
\end{subequations}
where $\mathbf{P}^{n-1}[i']$ is the optimal solution obtained from the subproblem at the $(n-1)$-th iteration, and $\partial \big\langle \|\mathbf{P}[i']\|_2^{n-1} \big\rangle$ represents the subgradient of the spectral norm at the $(n-1)$-th iteration. At this point, optimization (P7.3) becomes a convex optimization problem, which can be efficiently solved using the CVX toolbox. After the aforementioned linearization process, both the objective function and constraints of optimization (P7.3) are convex. Therefore, the problem has been transformed into a standard convex optimization problem, and the optimal solution satisfying the rank-1 constraint can be obtained by iteratively solving (P7.3) until the objective function value converges to 0. In actual numerical simulations, we usually set a sufficiently small convergence threshold $\epsilon$. When the condition is met, the iteration stops.

\begin{algorithm}[t]
\caption{MPC-Based Beamforming and Trajectory Optimization}
\label{alg:mpc_algorithm}
\begin{algorithmic}[1] 
\REQUIRE 
    Start/End points $\mathbf{q}_{\mathrm{A}}, \mathbf{q}_{\mathrm{B}}$, horizon $N_p$, thresholds $\epsilon_{in}, \epsilon_{out}$, max flight steps $T_{max}$, max AO iterations $I_{max}$. \\
    \textbf{Initialize:} $\mathbf{q}[0]\!=\!\mathbf{q}_{\mathrm{A}}, \mathbf{v}[0]\!=\!0$, and variables $\mathbf{W}_k, \mathbf{M}, \mathbf{\Theta}, \mathbf{\Psi}$.
\ENSURE 
    Trajectory $\mathbf{q}$, beamforming $\mathbf{W}_k$, AN $\mathbf{M}$, RIS phase $\mathbf{\Theta}, \mathbf{\Psi}$.

\STATE Set time slot $i \leftarrow 0$.
\WHILE{$||\mathbf{q}[i] - \mathbf{q}_{\mathrm{B}}|| > \epsilon_{out}$ \AND $i < T_{max}$} 
    \STATE \textbf{Step 1: Predict \& Initialize.} Generate guess for $i' \in \{i, \dots, i+N_p\}$, set AO index $\mu \leftarrow 0$.
    \STATE \textbf{Step 2: AO Loop.}
    \WHILE{$\mu < I_{max}$ \AND objective increase $> \epsilon_{in}$}
        \STATE Solve \textbf{(P4.1)}, update trajectory: $(\mathbf{q}^{(\mu+1)}, \mathbf{v}^{(\mu+1)}) \leftarrow (\mathbf{q}^{(\mu)}, \mathbf{v}^{(\mu)})$.
        \STATE Solve \textbf{(P5.2)}, update beamforming/AN: $(\mathbf{W}_k^{(\mu+1)}, \mathbf{M}^{(\mu+1)}) \leftarrow (\mathbf{W}_k^{(\mu)}, \mathbf{M}^{(\mu)})$.
        \STATE Solve \textbf{(P6.3)} and \textbf{(P7.3)}, update RIS phases: $\mathbf{\Theta}^{(\mu+1)} \leftarrow \mathbf{\Theta}^{(\mu)}$, $\mathbf{\Psi}^{(\mu+1)} \leftarrow \mathbf{\Psi}^{(\mu)}$.
        \STATE Update $\mu \leftarrow \mu + 1$ and check convergence.
    \ENDWHILE
    \STATE \textbf{Step 3: MPC Update.} Extract $\mathbf{v}^*[i]$, update state and parameters:
    \STATE $\mathbf{q}[i+1] \leftarrow \mathbf{q}[i] + \mathbf{v}^*[i] t_c + \mathbf{\Delta}_{dist}$ \textit{\% $\mathbf{\Delta}_{dist}$ denotes environmental perturbations}.
    \STATE Update real-time CSI, and set $i \leftarrow i + 1$.
\ENDWHILE
\RETURN Optimized variables for all time slots.
\end{algorithmic}
\end{algorithm}

\subsection{Convergence and Complexity Analysis}
\label{sec:analysis}
Based on the solutions to the sub-problems derived above, we propose the overall MPC-based beamforming and trajectory optimization algorithm. The detailed procedure is summarized in \textbf{Algorithm \ref{alg:mpc_algorithm}}. The following sections analyze the convergence and computational complexity of the proposed algorithm, as detailed below:
\subsubsection{Convergence Analysis}
Let $\mathcal{V}^{(\mu)} \triangleq \{\mathbf{q}^{(\mu)}, \allowbreak \mathbf{v}^{(\mu)}, \allowbreak \mathbf{W}_k^{(\mu)}, \allowbreak \mathbf{M}^{(\mu)}, \allowbreak \mathbf{\Theta}^{(\mu)}, \allowbreak \mathbf{\Psi}^{(\mu)}\}$ denote the optimization variable set at the $\mu$-th iteration, and let $\mathcal{O}(\mathcal{V}^{(\mu)})$ denote the generalized objective function of (P3), which incorporates the rank-1 penalty terms.

During the alternating optimization, updating each block sequentially guarantees a non-decreasing objective value. Specifically, the SCA-based trajectory update in (P4.1) leverages a first-order Taylor expansion that serves as a global lower bound, ensuring objective improvement. Furthermore, the exact optimizations for transceivers with AN in (P5.2) and the penalty minimizations for RIS phase shifts in (P6.3) and (P7.3) inherently maximize $\mathcal{O}$. Consequently, we obtain the following inequality chain:
\begin{align}
    \mathcal{O}(\mathcal{V}^{(\mu)}) &\le \mathcal{O}(\mathbf{q}^{(\mu+1)}, \mathbf{v}^{(\mu+1)}, \mathbf{W}_k^{(\mu)}, \mathbf{M}^{(\mu)}, \mathbf{\Theta}^{(\mu)}, \mathbf{\Psi}^{(\mu)}) \notag \\
    &\le \mathcal{O}(\mathbf{q}^{(\mu+1)}, \mathbf{v}^{(\mu+1)}, \mathbf{W}_k^{(\mu+1)}, \mathbf{M}^{(\mu+1)}, \mathbf{\Theta}^{(\mu)}, \mathbf{\Psi}^{(\mu)}) \notag \\
    &\le \mathcal{O}(\mathcal{V}^{(\mu+1)}). \label{eq:convergence}
\end{align}
The inequality chain \eqref{eq:convergence} demonstrates that $\mathcal{O}(\mathcal{V}^{(\mu)})$ is monotonically non-decreasing over iterations. Since the system's average secrecy rate is strictly upper-bounded by the finite transmit power budget $P_{\max}$, the theoretical convergence of the proposed iterative algorithm is rigorously guaranteed.
\subsubsection{Complexity Analysis} 
The per-slot computational complexity is dominated by solving sub-problems (P4.1), (P5.2), and (P6.3)-(P7.3) within the inner alternating optimization (AO) loop. Based on the standard complexity of the interior-point method, the overall computational complexity is approximated as: 
\begin{equation}
\mathcal{C}_{total} \approx \mathcal{O} \left( I_{ao} \left[ N_p^{3.5} + N_p(K+1)^{3.5} N^{3.5} + 2N_p M^{3.5} \right] \right),
\end{equation}
where $I_{ao}$ is the number of AO iterations. The parameters $N_p$, $K$, $N$, and $M$ denote the prediction horizon length, the number of legitimate users, the number of DFRC BS antennas, and the number of aerial RIS reflecting elements, respectively.
\section{Numerical Results}
In this section, numerical results are presented to evaluate the proposed MPC-based joint optimization algorithm, implemented via the MATLAB CVX toolbox. Unless otherwise specified, the default system parameters are set as follows. The time slot interval is $t_c = 1$ s. For the UAV kinematics, the weight, rotor area, air density, and drag coefficient are set to $W = 10$ kg, $S = 0.1$ m$^2$, $\rho = 1.225$ kg/m$^3$, and $\zeta = 0.08$, respectively. The maximum horizontal velocity, vertical velocity, and acceleration are bounded by $V_{\max} = 20$ m/s, $U_{\max} = 5$ m/s, and $a_{\max} = 5$ m/s$^2$, with altitude limits $Z \in [20, 60]$ m. The DFRC BS, aerial RIS, and legitimate users are equipped with $N_t = 8$, $M = 16$, and $K = 5$ antennas/elements, respectively. The path loss exponents and Rician factors for the BS-RIS, RIS-user, and RIS-eavesdropper links are $\alpha \in \{2.2, 2.5, 2.5\}$ and $\kappa \in \{10, 10, 10\}$, respectively. Finally, the maximum BS transmit power is $P_{\max} = 30$ dBm, and the receiver noise power is $\sigma_k^2 = \sigma_e^2 = -90$ dBm.

To comprehensively evaluate the proposed method, we consider five benchmarks: 1) \text{Traditional BCD}, which optimizes the trajectory offline without dynamic feedback; 2) \text{Fixed Trajectory}, where the UAV flies straight at a constant speed; 3) \text{Random RIS}, which sets a random RIS phase shift matrix; 4) \text{No AN}, which designs information beams without AN injection; and 5) \text{ZF Beamforming}, replacing complex joint optimization with low-complexity zero-forcing (ZF) precoding.

\begin{figure}[H]  
\centering
\includegraphics[width=0.75
\linewidth]{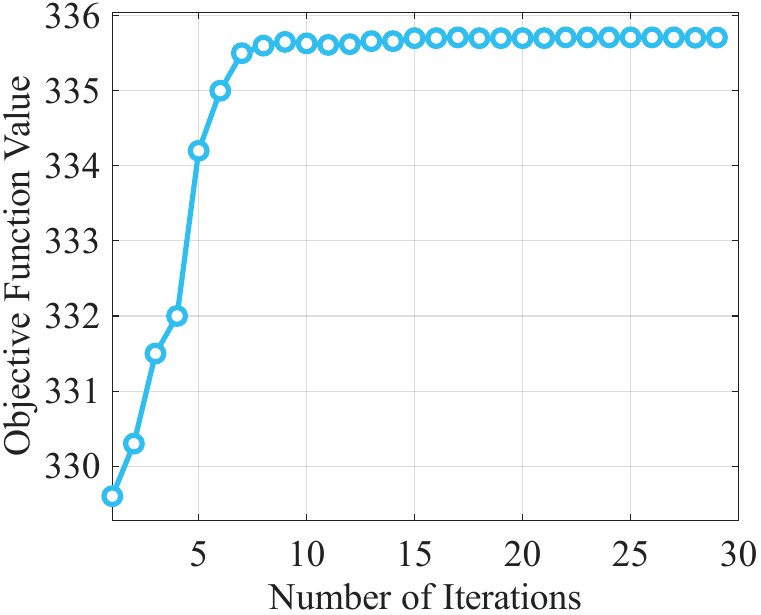} 
    \caption{\ Convergence behavior of the proposed algorithm.}
    \label{fig:2}
    \vspace{2em}
\end{figure}

\begin{figure}[H]  
\centering
\includegraphics[width=0.75\linewidth]{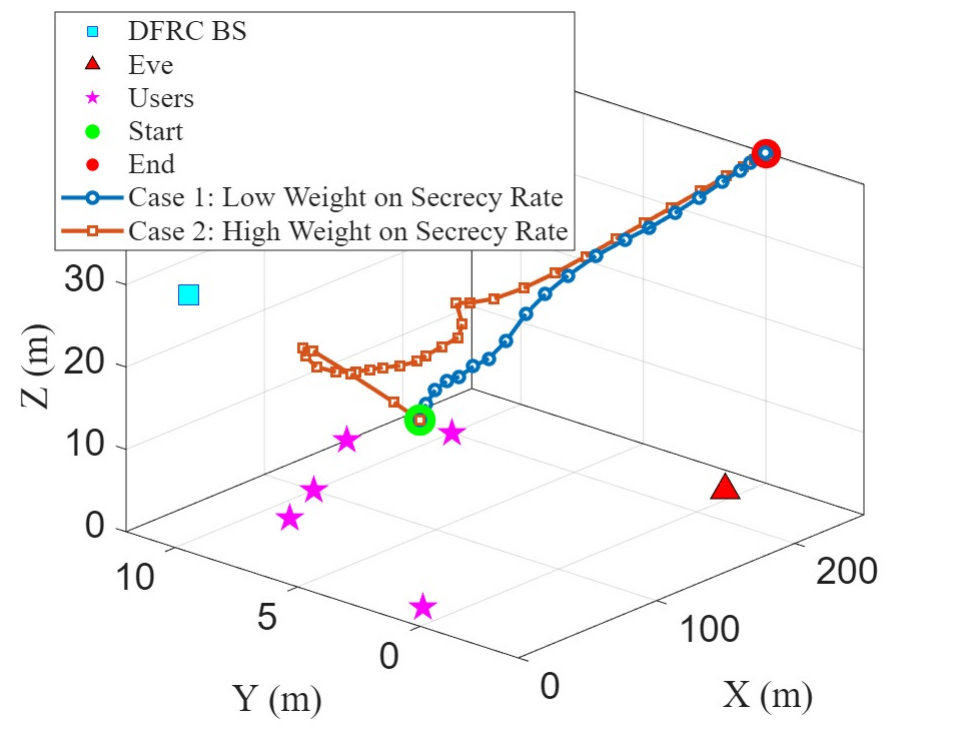} 
    \caption{\ UAV trajectories under different optimization objectives.}
    \label{fig:3}
    \vspace{2em}
\end{figure}

\begin{figure}[H]  
\centering
\includegraphics[width=0.75\linewidth]{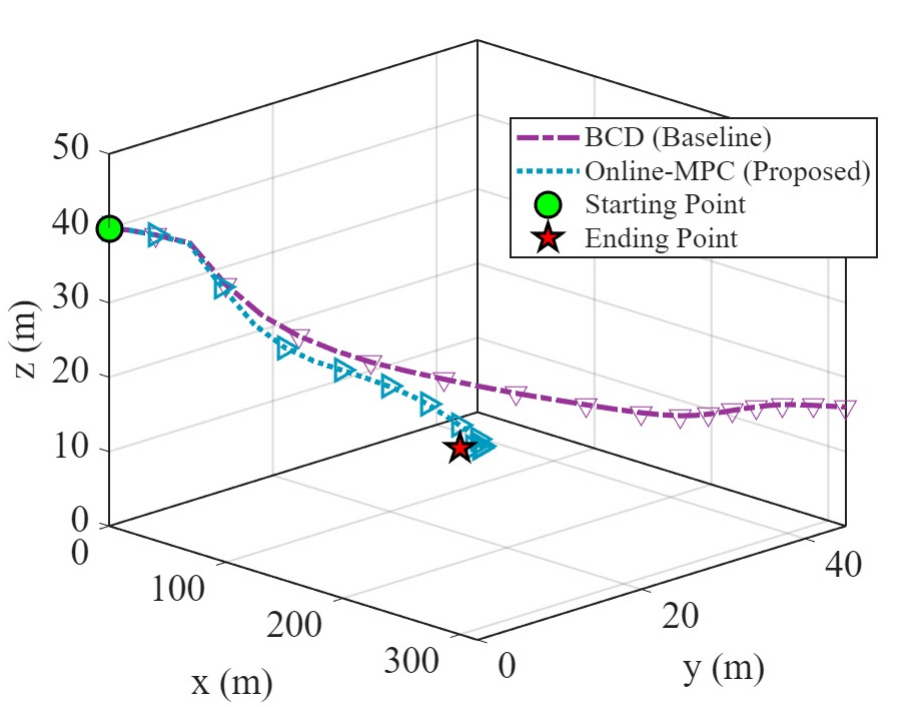} 
    \caption{\ UAV 3D trajectories under continuous interference.}
    \label{fig:4}
    \vspace{2em}
\end{figure}

Fig. \ref{fig:2} illustrates the convergence of the proposed algorithm. The curve indicates that as the iterations proceed, the objective function value increases steadily, and the gradient of the curve approaches zero. This trend implies that the objective function converges to a specific value, thereby demonstrating the convergence of the proposed algorithm.
In summary, the processes above solve the non-convexity of the objective function and all non-convex constraints. We successfully transform the optimization problem into a SDP problem that can be efficiently solved using standard convex optimization solvers.
\vspace{-0.5em}
\vspace{-0.5em}
Fig.~\ref{fig:3} compares two distinct UAV trajectory strategies in an interference-free scenario. When prioritizing energy and path length minimization (blue curve), the UAV adopts a direct, "high-altitude cruising" path. While this shortens flight time, the large distance to legitimate users incurs severe path loss, thereby degrading the communication rate. Conversely, when maximizing the secrecy rate (orange curve), the trajectory shifts significantly. The UAV actively lowers its altitude and deviates horizontally to approach the legitimate users, while simultaneously maneuvering to evade eavesdroppers. This effectively suppresses the interception rate and ensures high-quality secure communication.
\vspace{-0.5em}
\vspace{-0.5em}
Fig.~\ref{fig:4} compares the 3D flight trajectories of the traditional BCD and the proposed online MPC algorithms under continuous environmental interference. Lacking a real-time feedback mechanism, the BCD algorithm fails to correct deviations. Consequently, cumulative wind-induced errors cause the UAV to deviate severely and ultimately miss the destination. In contrast, the proposed MPC algorithm demonstrates superior robustness. By leveraging receding horizon optimization, it recalculates the optimal control strategy based on the actual position in each time slot. This real-time adjustment effectively compensates for continuous crosswind disturbances, ensuring the UAV reaches its destination precisely.

\begin{figure}[H]  
\centering
\includegraphics[width=0.75\linewidth]{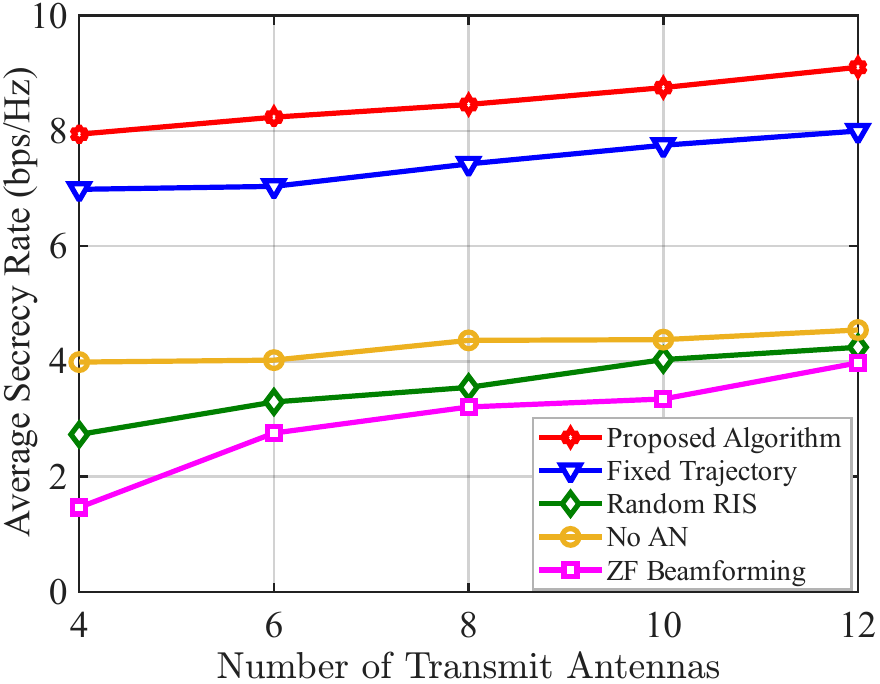} 
    \caption{\ Average secrecy rate vs. number of transmit antennas $N_t$.}
    \label{fig:5}
    \vspace{2em}
\end{figure}

\begin{figure}[H]  
\centering
\includegraphics[width=0.75\linewidth]{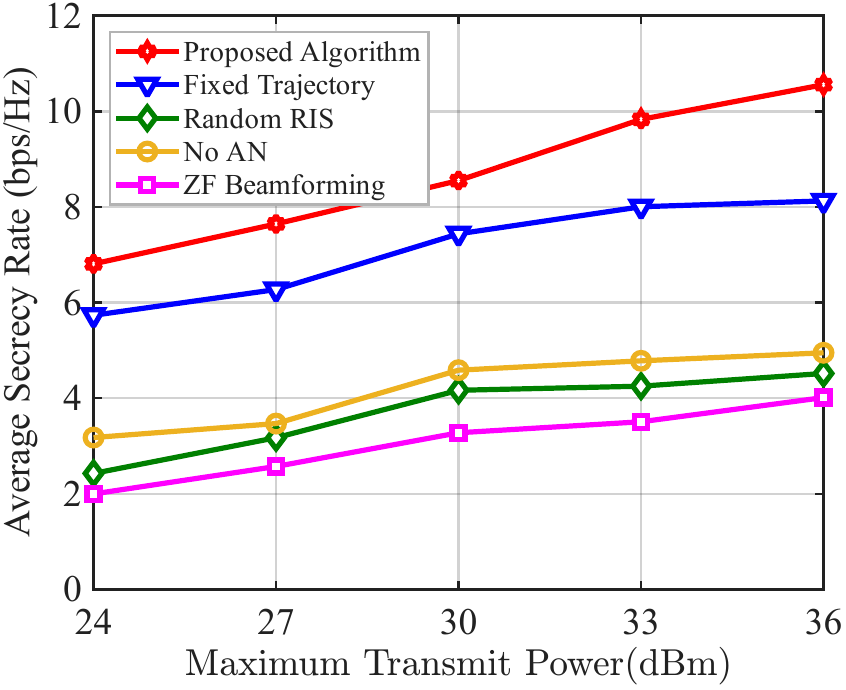} 
    \caption{\ Average secrecy rate vs. maximum transmit power $P_{\max}$.}
    \label{fig:6}
    \vspace{2em}
\end{figure}

\begin{figure}[H]  
\centering
\includegraphics[width=0.75\linewidth]{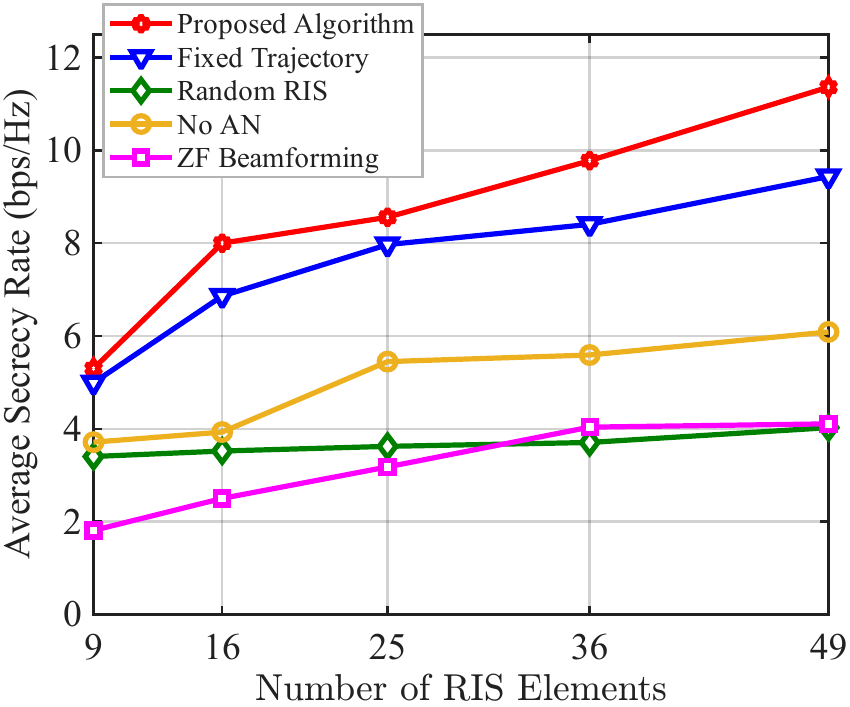}
    \caption{Average secrecy rate vs. number of reflecting elements $M$.}
    \label{fig:7}
    \vspace{2em}
\end{figure}

\begin{figure}[H]  
\centering
\includegraphics[width=0.75\linewidth]{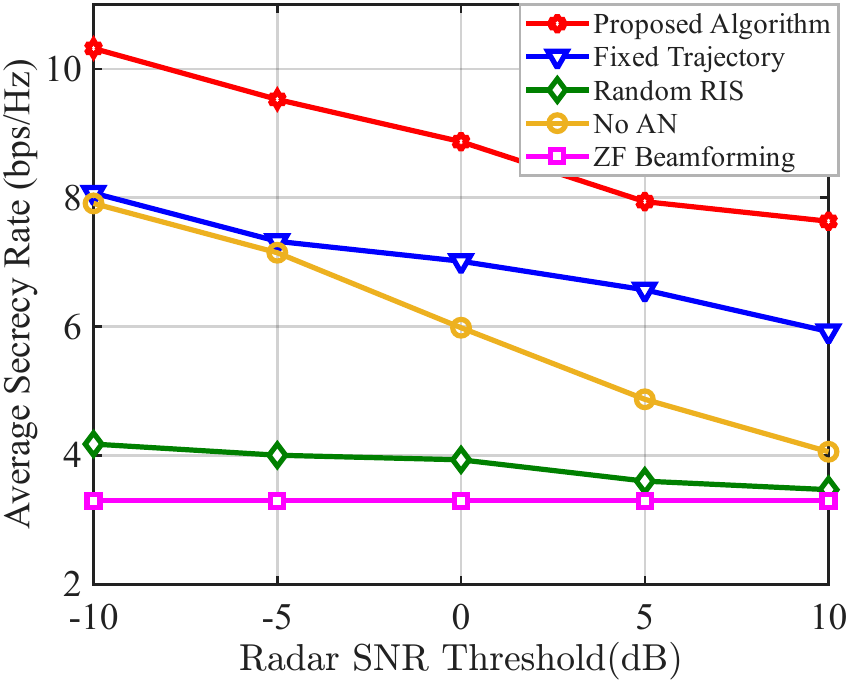} 
    \caption{\ Average secrecy rate vs. radar sensing SNR threshold $\gamma^r$.}
    \label{fig:8}
    \vspace{2em}
\end{figure}

Fig. \ref{fig:5} shows the variation of the average secrecy rate of legitimate users with the number of DFRC BS transmit antennas, $N_t$, under different schemes. It can be seen that as the number of antennas increases, the system gains higher spatial degrees of freedom and beamforming gain, resulting in a monotonic increasing trend in the achievable secrecy rate for all schemes. The proposed scheme exhibits optimal performance under all antenna configurations. Its core advantage lies in the deep joint optimization of the UAV trajectory, beamforming, and RIS phase shifts realized through the MPC framework.
\vspace{-0.5em}
\vspace{-0.3em}
\vspace{-0.5em}
\vspace{-0.2em}
Fig.~\ref{fig:6} plots the average secrecy rate versus the maximum transmit power, $P_{\max}$, at the DFRC BS. As $P_{\max}$ increases, the secrecy rate of all schemes grows monotonically. This enhancement stems from the expanded power budget, which provides the BS with greater beamforming flexibility. While strictly satisfying the radar sensing threshold, the BS can allocate stronger directional signals to legitimate users and inject higher-power AN, thereby boosting the system's secrecy capacity. The proposed scheme maintains a substantial performance lead over all baselines, verifying the superior efficacy of joint resource management in suppressing eavesdropping links and securing the communication environment.
\vspace{-0.5em}
\vspace{-0.5em}
Fig.~\ref{fig:7} illustrates the impact of the number of aerial RIS reflecting elements, $M$, on the average secrecy rate. For all schemes, the secrecy rate improves monotonically with $M$ due to the enhanced passive beamforming gain. Specifically, larger arrays provide more spatial degrees of freedom (DoFs) to precisely steer signals toward users while forming deep spatial nulls against eavesdroppers. The proposed MPC algorithm consistently outperforms static baselines (e.g., Fixed Trajectory and Random RIS) by dynamically co-optimizing the 3D trajectory and high-dimensional phase shifts, thereby maximizing the utilization of these spatial DoFs to secure communications.
\vspace{-0.5em} 
\vspace{-0.3em} 
Fig.~\ref{fig:8} depicts the average secrecy rate versus the radar sensing SNR threshold, $\gamma^r$. The ubiquitous downward trend across all schemes highlights the fundamental sensing-security trade-off in ISAC systems. Specifically, stringent sensing demands force the BS to allocate more power and spatial DoFs to sensing beams, thereby depleting resources for legitimate signal enhancement and AN injection. Despite this, the proposed algorithm consistently achieves the highest secrecy rate, demonstrating the superiority of receding horizon optimization in dynamically scheduling resources to maximize security under strict sensing constraints. Furthermore, the sharpest performance drop observed in the No AN scheme validates the critical role of AN in safeguarding communications within resource-limited environments.
\vspace{-0.5em}
\vspace{-0.3em}
\section{Conclusion}\label{sec5}
In this paper, we studied the adaptive secure ISAC design for aerial RIS-enabled systems in dynamic environments. By integrating UAV-mounted RIS deployment with predictive control, we developed an MPC-based online optimization framework to enable real-time adjustment of communication and sensing strategies. The proposed framework effectively addresses the challenges of long-term decision-making and dynamic uncertainties by optimizing the UAV trajectory, BS beamforming with AN, and RIS phase configuration in a receding-horizon manner. An efficient alternating optimization approach was further designed to handle the coupled non-convex optimization problem. Simulation results verified that the proposed scheme achieves substantial secrecy performance improvements over conventional approaches, while maintaining reliable sensing performance and exhibiting enhanced adaptability to environmental variations. These results highlight the potential of aerial RIS-assisted secure ISAC systems for future dynamic and intelligent wireless networks.
\vspace{1.0em}
\balance
\bibliographystyle{IEEEtran}
\bibliography{reference}
\vspace{1.0em}

	\end{document}